\documentclass[sigconf,balance=false]{acmart}
\usepackage{popets}
\setcopyright{popets}
\copyrightyear{YYYY}
\acmYear{YYYY}
\acmVolume{YYYY}
\acmNumber{X}
\acmDOI{XXXXXXX.XXXXXXX}
\acmISBN{}
\acmConference{Proceedings on Privacy Enhancing Technologies}
\usepackage{tikz}
\usetikzlibrary{positioning, calc}
\usepackage{
  color,
  float,
  epsfig,
  wrapfig,
  graphics,
  graphicx,
  subcaption
}

\usepackage{nicefrac}
\usepackage{siunitx}
\usepackage{array,framed}
\usepackage{booktabs}
\usepackage{subcaption}
\usepackage{stackengine}
\usepackage[most]{tcolorbox}

\usepackage{textcomp}
\usepackage{tikz}
\usepackage{setspace}
\usepackage{colortbl}
\usepackage{amsfonts}

\usepackage{bbm}
\usepackage{amsmath}
\usepackage{algorithm}
\usepackage{algorithmic}
\usepackage{graphics}
\usepackage{xparse} 
\usepackage{xspace}
\usepackage{multirow}
\usepackage{csvsimple}
\usepackage{balance}
\usepackage{xcolor}
\usepackage{flushend}
\definecolor{takeaways}{HTML}{F5F5F5}
\definecolor{ref}{HTML}{116D6E}
\definecolor{cite}{HTML}{E55807}

\newcommand{\myattackone}{\ensuremath{\mathsf{GSA_1}}\xspace}
\newcommand{\myattacktwo}{\ensuremath{\mathsf{GSA_2}}\xspace}

\usepackage{xcolor}
\definecolor{shadow_member}{rgb}{0.4 0.76078431 0.64705882}
\definecolor{shadow_non-member}{rgb}{0.55294118 0.62745098 0.79607843}
\definecolor{target_member}{rgb}{0.98823529 0.55294118 0.38431373}
\definecolor{target_non-member}{rgb}{0.90588235 0.54117647 0.76470588}
\usepackage{mathtools,}
\usepackage{cancel}
\usepackage{dsfont}
\usepackage{float}

\usepackage{
  tikz,
  pgfplots,
  pgfplotstable
}
\usepackage{hyperref}
\usepackage{enumitem}

\usetikzlibrary{
  shapes.geometric,
  arrows,
  external,
  pgfplots.groupplots,
  matrix
}

\usepackage{xparse}
\newcommand{\bnm}{\begin{newmath}}
\newcommand{\enm}{\end{newmath}}

\newcommand{\bea}{\begin{eqnarray*}}%
\newcommand{\eea}{\end{eqnarray*}}%

\newcommand{\bne}{\begin{newequation}}
\newcommand{\ene}{\end{newequation}}

\newcommand{\bal}{\begin{newalign}}
\newcommand{\eal}{\end{newalign}}

\newenvironment{newalign}{\begin{align}%
\setlength{\abovedisplayskip}{4pt}%
\setlength{\belowdisplayskip}{4pt}%
\setlength{\abovedisplayshortskip}{6pt}%
\setlength{\belowdisplayshortskip}{6pt} }{\end{align}}

\newenvironment{newmath}{\begin{displaymath}%
\setlength{\abovedisplayskip}{4pt}%
\setlength{\belowdisplayskip}{4pt}%
\setlength{\abovedisplayshortskip}{6pt}%
\setlength{\belowdisplayshortskip}{6pt} }{\end{displaymath}}

\newenvironment{newequation}{\begin{equation}%
\setlength{\abovedisplayskip}{4pt}%
\setlength{\belowdisplayskip}{4pt}%
\setlength{\abovedisplayshortskip}{6pt}%
\setlength{\belowdisplayshortskip}{6pt} }{\end{equation}}

\newcounter{ctr}

\newcounter{mytable}
\def\mytable{\begin{centering}\refstepcounter{mytable}}
\def\endmytable{\end{centering}}

\newcounter{myfig}
\def\myfig{\begin{centering}\refstepcounter{myfig}}
\def\endmyfig{\end{centering}}

\newlength{\saveparindent}
\newlength{\saveparskip}
\newcommand{\E}{\mathbb{E}}

\renewcommand{\eqref}[1]{\mbox{Equation~(\ref{#1})}}

\def \part {part}

\renewcommand{\paragraph}[1]{\vspace*{6pt}\noindent\textbf{#1}\;}

\def \blackslug{\hbox{\hskip 1pt \vrule width 4pt height 8pt
    depth 1.5pt \hskip 1pt}}
\def \qed{\quad\blackslug\lower 8.5pt\null\par}

\newcounter{mynote}[section]

\newcommand\ignore[1]{}

\newcounter{rcnote}[section]

\newcounter{mrnote}[section]

\newcounter{fknote}[section]

\newcounter{anote}[section]

\DeclareMathSymbol{\mlq}{\mathord}{operators}{``}
\DeclareMathSymbol{\mrq}{\mathord}{operators}{`'}

\newcommand{\rhf}[2]{R_{f, \gamma}}

\DeclareDocumentCommand{\edist}{o o}{
  \ensuremath{
    \IfNoValueTF{#1}{{d}}{{\sf d}(#1,#2)}
  }
}

\newcommand{\olrk}[1]{\ifx\nursymbol#1\else\!\!\mskip4.5mu plus 0.5mu\left(\mskip0.5mu plus0.5mu #1\mskip1.5mu plus0.5mu \right)\fi}

\NewDocumentCommand{\indseq}{ O{1} O{r} }{{#1}\ldots {#2}}

\newcommand{\graysquare}{
    \begin{tikzpicture}
    \fill[gray] (0,0) rectangle (0.25,0.25); 
    \end{tikzpicture}
}

\usepackage{xspace,comment}

\begin{document}

\title[White-box Membership Inference Attacks against Diffusion Models]{White-box Membership Inference Attacks against Diffusion Models}


\author{Yan Pang}
\affiliation{%
  \institution{University of Virginia}
  \city{}
  \state{}
  \country{}
  }
\email{yanpang@virginia.edu}

\author{Tianhao Wang}
\affiliation{%
  \institution{University of Virginia}
  \city{}
  \state{}
  \country{}
  }
\email{tianhao@virginia.edu}

\author{Xuhui Kang}
\affiliation{%
  \institution{University of Virginia}
  \city{}
  \state{}
  \country{}
  }
\email{qhv6ku@virginia.edu}

\author{Mengdi Huai}
\affiliation{%
 \institution{Iowa State University}
  \city{}
  \state{}
  \country{}
 }
\email{mdhuai@iastate.edu}

\author{Yang Zhang}
\affiliation{%
  \institution{CISPA Helmholtz Center for Information Security}
  \city{}
  \state{}
  \country{}
  }
\email{zhang@cispa.de}









\begin{abstract}
Diffusion models have begun to overshadow GANs and other generative models in industrial applications due to their superior image generation performance. 
The complex architecture of these models furnishes an extensive array of attack features.  
In light of this, we aim to design membership inference attacks (MIAs) catered to diffusion models. We first conduct an exhaustive analysis of existing MIAs on diffusion models, taking into account factors such as black-box/white-box models and the selection of attack features. 
We found that white-box attacks are highly applicable in real-world scenarios, and the most effective attacks presently are white-box.
Departing from earlier research, which employs model loss as the attack feature for white-box MIAs, we employ model gradients in our attack, leveraging the fact that these gradients provide a more profound understanding of model responses to various samples. We subject these models to rigorous testing across a range of parameters, including training steps, timestep sampling frequency, diffusion steps, and data variance. Across all experimental settings, our method consistently demonstrated near-flawless attack performance, with attack success rate approaching $100\%$ and attack AUCROC near $1.0$. We also evaluated our attack against common defense mechanisms, and observed our attacks continue to exhibit commendable performance.
We provide access to our code\footnote{\url{https://github.com/py85252876/GSA}}.
\end{abstract}

\maketitle

\section{Introduction}
Recently, diffusion models have gained significant attention, and various applications are emerging. These models~\cite{ho2020denoising,nichol2021glide,ramesh2022hierarchical,ramesh2021zero,saharia2022photorealistic,sohl2015deep,song2019generative} rely on a progressive denoising process to generate images, resulting in improved image quality compared to previous models like GANs~\cite{creswell2018generative,dhariwal2021diffusion} and VAEs~\cite{liang2018variational}. 
%
Leading models primarily fall into two categories. The first category encompasses diffusion-based architectures such as GLIDE~\cite{nichol2021glide}, Stable Diffusion model~\cite{rombach2022highresolution}, DALL-E 2 \cite{ramesh2022hierarchical}, and Imagen \cite{saharia2022photorealistic}. 
The second category comprises representative sequence-to-sequence models like DALL-E~\cite{ramesh2021zero}, Parti~\cite{yu2022scaling}, and CogView~\cite{ding2021cogview}. Current text-to-image models possess the capability to generate exquisite and intricately detailed images based on textual inputs, finding extensive applications across various domains such as graphic design and illustration. 
%
%
%
%
While diffusion models can be employed to synthesize distinct artistic styles, they often necessitate training on extensive sets of sensitive data. 
Thus, investigating membership inference attacks (MIAs)~\cite{shokri2017membership}, which aim to determine whether specific samples are present in the diffusion model's training data, is of paramount importance.

Numerous studies have been conducted on classification models~\cite{carlini2022membership,li2021membership,liu2022membership,salem2018ml,shokri2017membership,ye2022enhanced,yeom2018privacy}, GANs~\cite{chen2020gan,hayes2017logan,hilprecht2019monte,hu2021membership,mukherjee2021privgan}, and others. However, due to the unique training and inference method of diffusion models, previous attack methods~\cite{yeom2018privacy} are no longer suitable. For instance, in classification models, the model's final output is generally used as the attack feature, relying on the model's overfitting to the training data, which leads to differences in classification confidence. Additionally, previous work on generative models such as GANs focused on utilizing the discriminator for determination~\cite{mukherjee2021privgan}. Since the diffusion model does not have a discriminator, which makes it different from GANs, a new attack method must be specifically designed for diffusion models.

Some preliminary efforts have been devoted to conducting MIA on diffusion models~\cite{hu2023membership,matsumoto2023membership,carlini2023extracting}. 
However, it merits our attention that these investigations, akin to many others in this domain, predominantly concentrate on loss- and threshold-based attacks. We postulate that different layers in a neural network learn distinct features and, therefore, store varying amounts of information~\cite{yosinski2014transferable}. Evaluations based solely on loss could potentially overlook substantial information~\cite{mukherjee2021privgan}. Consequently, a more comprehensive perspective of the model's response to a sample could be attained by considering gradient information from each layer post-backpropagation in addition to the loss incurred by the model.

The main challenges of utilizing gradients for MIAs are the excessive computation overhead and the overfitting issue of training the attack model (given the large size of diffusion models, gradients could have millions of dimensions).  We carefully analyze ways to reduce dimensionality and propose a framework incorporating subsampling and aggregation.  We call our framework Gradient attack based on Subsampling and Aggregation (GSA) and initiate two instances, \myattackone and \myattacktwo, demonstrating different trade-offs within the GSA framework.  

To ensure the comprehensiveness and integrity of our investigation, we conduct experiments on the fundamental unconditional Denoising Diffusion Probabilistic Models (DDPM)~\cite{ho2020denoising} and the state-of-the-art Imagen model~\cite{saharia2022photorealistic}, which presently leads the text-to-image domain. CIFAR-10 and ImageNet datasets are utilized to train the unconditional diffusion models, while the MS COCO dataset is employed to train the Imagen model. We further explore the influence of varying parameters on the effectiveness of the attack. Ultimately, we validate the effectiveness of our attack strategy with a near 100$\%$ success rate, thus underscoring the imperative need for addressing the security aspects of diffusion models.

The contributions of our work are two-fold:


\begin{itemize}[leftmargin=*]
    \item  We have analyzed membership inference attacks on diffusion models in existing research. Moreover, we have conceptualized our attack for new practical scenarios and conducted analyses across various dimensions, such as timesteps and model layers.
    \item We conducted experiments on three datasets using the traditional DDPM model and the cutting-edge text-to-image model, Imagen. Our results demonstrate extremely high accuracy across four evaluation metrics, underscoring the effectiveness of using gradients as attack features.
\end{itemize}


\paragraph{Roadmap.}
In~\autoref{background}, we introduce the background of diffusion models and delve into membership inference attacks. We also discuss the challenges we encountered and review existing attacks on diffusion models. In~\autoref{methodology}, we present our attack strategy. The experimental setup is detailed in~\autoref{experimental_setup}, while~\autoref{evaluation_results} showcases the results of these experiments. In~\autoref{Ablation}, we apply our GSA framework at the model layer level, demonstrating a further reduction in computational time. ~\autoref{defense} illustrates the performance of our attack under various defense strategies. The limitations of our attack are discussed in~\autoref{limitation}.~\autoref{related_work} touches upon related works, and finally, we conclude in \autoref{conclusion}.

\section{Background}\label{background}
\subsection{Diffusion Models}
\smallskip\noindent

The work of the Denoising Diffusion Probabilistic Models~\cite{ho2020denoising} (DDPM) has drawn considerable attention and led to the recent development of diffusion models~\cite{sohl2015deep, song2019generative}, which are characteristically described as ``progressively denoising to obtain the true image''. There are two categories of diffusion models: unconditional diffusion models, which do not incorporate any guiding input for image output, and conditional diffusion models, which were developed subsequently and generate images based on provided inputs information, such as labels~\cite{dhariwal2021diffusion,ho2022cascaded}, text~\cite{nichol2021glide,ramesh2022hierarchical,saharia2022photorealistic,rombach2022highresolution,ho2022classifierfree}, or low-resolution images~\cite{saharia2022palette,saharia2022image}.

\paragraph{Unconditional Diffusion Models.}
A diffusion model has two phases. First, during the forward process, the model progressively adds standard Gaussian noise to the true image $x_0$ through $T$ steps. The image at time $t$ is given by 
\begin{align}
    x_t = \sqrt{\bar{\alpha}_t} x_{0} + \sqrt{1 - \bar{\alpha}_t} \epsilon_t\label{eq:x_t}
\end{align} where $\epsilon_t$ represents the standard Gaussian noise obtained from the reparameterization trick. Furthermore, $\bar{\alpha}_t$ is defined as the product $\prod_{i=1}^{t} \alpha_i,$ with each parameter $\alpha_i$ monotonically decreasing and lying in the interval $[0,1]$.




Second, the reverse process begins with the noise image $x'_{T}$, where $x'_T \sim \mathcal{N} (0, I)$, and it progressively denoises to yield $x'_{T-1}, x'_{T-2}$, $\ldots$, $x'_0$ through the neural network (e.g., U-Net) $\epsilon_{\theta}$, parameterized by $\theta$. Specifically, $\epsilon_{\theta}$ takes a image $x'_t$ and a timestep $t$ as inputs, and predicts the noise, represented by $\epsilon_{\theta}(x'_t, t)$ that should be eliminated at step $t$. The final goal is to maximize the similarity between each pair of original image $x_{0}$ and the denoised image $x'_{0}$. 

During the training phase, the objective is to minimize the loss, which is defined as the expected squared $\ell_2$ error. This error is evaluated overall $\epsilon_t$ and the training sample $x_0$, as given by:
\begin{align}
    L_t(\theta) = \mathbb{E}_{x_0,\epsilon_t} \left[ \lVert \epsilon_t - \epsilon_{\theta}(\sqrt{\bar{\alpha}_t} x_0 + \sqrt{1 - \bar{\alpha}_t} \epsilon_t, t) \rVert^2_2\right]\;.
    \label{eq:diffusion-loss}
\end{align} 

More details can be found in \hyperref[appendix:DDPM]{Appendix~\ref*{appendix:DDPM}}.

\paragraph{Conditional Diffusion Models.}
As the study of diffusion models deepens, it has been discovered that classifiers can be utilized to guide the diffusion model generation~\cite{dhariwal2021diffusion}. Specifically, given a pre-trained classifier $M$ and a target class $c$, one can derive `directional information', $\nabla_{x_t}\log M(x_t|y)$, for an image $x_t$ and fuse it to the generation process of unconditional diffusion models.
%
%

In the text domain, Imagen employs T5, a significant language model~\cite{raffel2020exploring}, as a text encoder to guide the generation process through text embeddings~\cite{saharia2022photorealistic}.
Specifically, a distinct time embedding vector is constructed and modified during each timestep to align with the image's dimensions. The text embedding extracted from T5 is then incorporated with the time embedding and image to generate the conditional image.

\subsection{Membership Inference Attack}
\smallskip\noindent
Membership inference attack (MIA) tries to predict if a given sample was part of the training set used to train the target model. 
It has been widely applied to different deep learning models, including classification 
models~\cite{carlini2022membership,li2021membership,liu2022membership,salem2018ml,shokri2017membership,ye2022enhanced,yeom2018privacy}, generative adversarial networks 
(GANs)~\cite{chen2020gan,hayes2017logan,hilprecht2019monte,hu2021membership,mukherjee2021privgan}, and diffusion models~\cite{wu2022membership,hu2023membership, carlini2023extracting,duan2023diffusion,kong2023efficient,matsumoto2023membership}. MIA exploits the differential responses exhibited by machine learning models to training data. Specifically, these models react differently to samples they have been trained on, termed `member samples', versus unfamiliar `non-member samples'.

Shokri et al.~\cite{shokri2017membership} first proposed the technique of {shadow training}. This involves training shadow models to imitate the behavior of the target model. 
%
%
An attack model is then trained based on the output of the shadow models. This transforms membership inference into a classification problem.

%
%

Considering the increased computational overhead of training a machine learning model as an attack model, Yeom et al.~\cite{yeom2018privacy} proposed a more streamlined and resource-efficient approach—the {\it threshold-based} MIA. This method begins with the computation of loss values from the model's output prediction vector. These calculated loss metrics are subsequently compared against a chosen threshold to infer the membership status of a data record.

Carlini et al.~\cite{carlini2022membership} argue that while {threshold-based attacks} are effective for non-membership inference, they lack precision for member sample classification. This discrepancy arises as the approach simplifies the comparison process by scaling all samples based on their loss values, potentially omitting crucial sample-specific properties. To address this, Carlini et al. propose an alternative approach called Likelihood Ratio Attack (LiRA), which derives two distributions from the model's confidence values. These distributions are then used to determine the membership status of a given sample, thereby offering a more balanced evaluation of both member and non-member sets.

\subsection{Problem Formulation} 
\smallskip\noindent
In this paper, we investigate MIA in diffusion models. 
We are given a target model. 
The task is to predict whether a certain sample is part of the training dataset.
MIA on diffusion models (compared to classifiers) presents distinct challenges:
Classic classifier models yield vectors. Thus people can use its prediction vector as a feature for MIA~\cite{carlini2022membership,liu2022membership,salem2018ml,shokri2017membership,ye2022enhanced,yeom2018privacy}, which constitutes a black-box attack. 
Diffusion models produce images as outputs, making it challenging to launch an attack on a diffusion model using only its output, i.e., the image. The current state-of-the-art attacks on diffusion models are predominantly white-box, relying on the loss generated during the evaluation process, as noted in~\autoref{tab:existing_methods}. Our work is mainly focused on exploring how to get effective attack features. After getting the attack features (gradient data), we use it to train a machine learning model (i.e., XGBoost, MLP) as the attack model to identify the data sample.  
\smallskip\noindent

\noindent\textbf{Threat Model.} We operate under the assumption that an attacker possesses white-box access to the target model, encompassing its architectural intricacies and specific parameter details. In the context of conditional diffusion models, we assume that the attacker knows all modalities (for instance, image-text pairs) pertaining to the victim models. The same assumption has also been adopted in several existing works, which we will discuss in detail later~\cite{carlini2023extracting,hu2023membership,matsumoto2023membership}.
As more people openly share their model architectures and pre-trained checkpoints (like in HuggingFace\footnote{A multitude of model cards can be found on the Hugging Face website. \url{https://huggingface.co/models}}), the scenario is realistic. 
A motivating example is an artist who suspects his artwork is being used without permission to train a diffusion model. This model is subsequently uploaded to the HuggingFace website. As a result, others can use it to generate images that mimic the artist's unique style. Clearly, this constitutes a severe violation of the artist's intellectual property rights. The artist, as a result, downloads the model and checks whether their artwork is used to train the model. 
\subsection{Existing Work}

\begin{table}[t]
    \centering
    \Large
    \caption{
    Compared with existing work, we argue that with white-box access, using gradients is more effective. We also evaluated more comprehensively on larger datasets.
    }
    \label{tab:existing_methods}
    \resizebox{0.47\textwidth}{!}{
    \begin{tabular}{cccc}
         \toprule[1.3pt]
         \multirow{2}{*}{\textbf{\begin{tabular}{@{}c@{}}Attack \end{tabular}}} &  \multirow{2}{*}{\textbf{\begin{tabular}{@{}c@{}}Feature \end{tabular}}} &  \multirow{2}{*}{\textbf{\begin{tabular}{@{}c@{}}Victim \\ target \end{tabular}}} & \multirow{2}{*}{\textbf{\begin{tabular}{@{}c@{}}Training \\ dataset \end{tabular}}} \\ \\ 
         \midrule[0.5pt]
         \multirow{2}{*}{\begin{tabular}{@{}c@{}}~\cite{carlini2023extracting} \end{tabular}} & \multirow{2}{*}{\begin{tabular}{@{}c@{}}Loss (LiRA)\end{tabular}} & \multirow{2}{*}{\begin{tabular}{c}Unconditional \\ Conditional \end{tabular}} & \multirow{2}{*}{\begin{tabular}{@{}c@{}}CIFAR-10 \end{tabular}} \\ \\ 
         \multirow{2}{*}{\begin{tabular}{@{}c@{}}~\cite{hu2023membership} \end{tabular}} & \multirow{2}{*}{\begin{tabular}{@{}c@{}}Loss (Threshold) \end{tabular}} & \multirow{2}{*}{\begin{tabular}{c}Unconditional \end{tabular}} & \multirow{2}{*}{\begin{tabular}{@{}c@{}}FFHQ \\ DRD \end{tabular}} \\ \\ 
         \multirow{2}{*}{\begin{tabular}{@{}c@{}}~\cite{matsumoto2023membership} \end{tabular}} & \multirow{2}{*}{\begin{tabular}{@{}c@{}}Loss (Threshold) \end{tabular}} & \multirow{2}{*}{\begin{tabular}{c}Unconditional \end{tabular}} & \multirow{2}{*}{\begin{tabular}{@{}c@{}}CIFAR-10 \\ CelebA \end{tabular}} \\ \\ 
         \multirow{2}{*}{\begin{tabular}{@{}c@{}}Ours \end{tabular}} & \multirow{2}{*}{\begin{tabular}{@{}c@{}}Gradient (ML model)\end{tabular}} & \multirow{2}{*}{\begin{tabular}{c}Unconditional \\ Conditional \end{tabular}} & \multirow{2}{*}{\begin{tabular}{@{}c@{}}CIFAR-10 ImageNet \\ MS COCO \end{tabular}} \\ \\ \bottomrule[1.3pt]
    \end{tabular}}
    
\end{table}
\paragraph{Existing White-Box Attacks to Diffusion Models.} 
A key challenge in applying MIA is selecting the appropriate information/features to distinguish member and non-member samples. 
Most effective attacks on diffusion models predominantly employ white-box techniques~\cite{carlini2023extracting,hu2023membership,matsumoto2023membership}.

Hu et al.~\cite{hu2023membership} and Matsumoto et al.~\cite{matsumoto2023membership} suggested utilizing the loss, as defined in~\autoref{eq:diffusion-loss}, at each timestep $t$ as a feature in conjunction with a threshold-based MIA. Leveraging the loss directly as an attack vector presents the most intuitive attack approach. {However, the loss value differences between member and non-member samples vary across different timesteps. For each model, additional computation is required to identify the most effective range of timesteps, which greatly increases pre-computational cost and becomes impractical. Additionally, since the loss value is a scalar, it may lead to unstable attack accuracy due to insufficient information for reliable differentiation. In contrast, gradient data can effectively differentiate between member and non-member samples without requiring prior timestep selection. As high-dimensional data, it also enhances the accuracy and robustness of the attack.}

Carlini et al.~\cite{carlini2023extracting} also opted to employ loss and use the LiRA framework. 
%
In the context of the LiRA online framework, the attack strategy necessitates utilizing target points for the training of several shadow models, a process that is both computationally demanding and time-intensive. Subsequently, it constructs the \(\mathbb{D}_{\text{in}}\) and \(\mathbb{D}_{\text{out}}\) distributions at each timestep. In the original experiments reported in the paper, $16$ shadow models were trained to generate distributions for each timestep. For more sophisticated models, such as Stable Diffusion~\cite{rombach2022highresolution}, retraining a large cohort of shadow models to generate loss distributions poses a considerable challenge. In our work, we aim to use fewer shadow models to execute the attack while maintaining effectiveness and efficiency.
More details about LiRA can be found at~\hyperref[appendix:carlini]{Appendix~\ref*{appendix:carlini}}.

\paragraph{Other Attacks.}
Several studies have utilized the properties of DDIM~\cite{kim2022diffusionclip,song2020denoising,song2020score} (as detailed in \hyperref[appendix:DDPM]{Appendix~\ref*{appendix:DDPM}}) for attacks~\cite{duan2023diffusion,kong2023efficient}. However, these attacks are contingent on the deterministic reverse process of DDIM, and cannot be directly applied to DDPM. Detailed discussions of these attacks are deferred to \hyperref[appendix:secmi]{Appendix~\ref*{appendix:secmi}} and \hyperref[appendix:pia]{Appendix~\ref*{appendix:pia}}. 

Prior to diffusion models, there are also MIAs for GANs~\cite{chen2020gan,hayes2017logan,hilprecht2019monte,hu2021membership,mukherjee2021privgan}.
Note that GANs and diffusion models differ in their overall architecture; therefore, white-box attacks toward GANs are not directly applicable to diffusion models.
On the other hand, black-box attacks share similarities as both GANs and diffusion models are generative models.
In particular, inspired by the attacks of GAN-Leaks~\cite{chen2020gan}, Matsumoto et al.~\cite{matsumoto2023membership} proposed an attack that is based on the reconstruction error of the target image and a set of generated samples. We will present the details at \hyperref[appendix:ganleaks]{Appendix~\ref*{appendix:ganleaks}}, but the attack shows limited effectiveness. 

Meanwhile, Wu et al.~\cite{wu2022membership} carried out black-box attacks on pre-trained text-to-image diffusion models, launching attacks at both the pixel-level and semantic-level. However, their method does not employ the shadow model technique as proposed in~\cite{shokri2017membership}, instead conducting all experiments directly on the target model and selecting the training set of the pre-trained model as the member set. Consequently, this attack strategy is not universally effective for every victim model.

Hu et al.~\cite{hu2023membership} also initiated an interesting threat model (the so-called grey-box model or query-based model) where the attacker sees the intermediate denoised images and proposes an attack based on the similarities (likelihood) between pairs of these intermediate samples and those in the forward pass (details also in \hyperref[appendix:likelihood]{Appendix~\ref*{appendix:likelihood}}). 



%


\section{Methodology}
\label{methodology}
Previous works on attacking the diffusion model encompass black-box~\cite{matsumoto2023membership,wu2022membership}, gray-box~\cite{duan2023diffusion,hu2023membership,kong2023efficient}, and white-box approaches~\cite{carlini2023extracting,hu2023membership,matsumoto2023membership}. Upon comparing their accuracy and considering practical implications, we contend that white-box attacks on diffusion models are the most effective. 

\subsection{Theoretical Foundation and Challenge}
Current white-box attacks often manipulate the loss at different timesteps through various methods (e.g., threshold~\cite{hu2023membership,matsumoto2023membership} or distribution~\cite{carlini2023extracting}). However, it often necessitates a substantial amount of time to identify the timestep where the loss can most distinctly differentiate between the member and non-member set samples.
We argue that {\it rather than relying on the loss information, given white-box access, it could be more insightful to leverage gradient information} that better reflects the model's different responses to member samples and non-member samples. 
%
The intuition using gradients is, as gradients are generally very high-dimensional (than losses), it offers a more nuanced representation of its response to an input target point compared to mere loss values. 

\begin{figure}[!t]
    \centering
    \includegraphics[width = 3.4in]{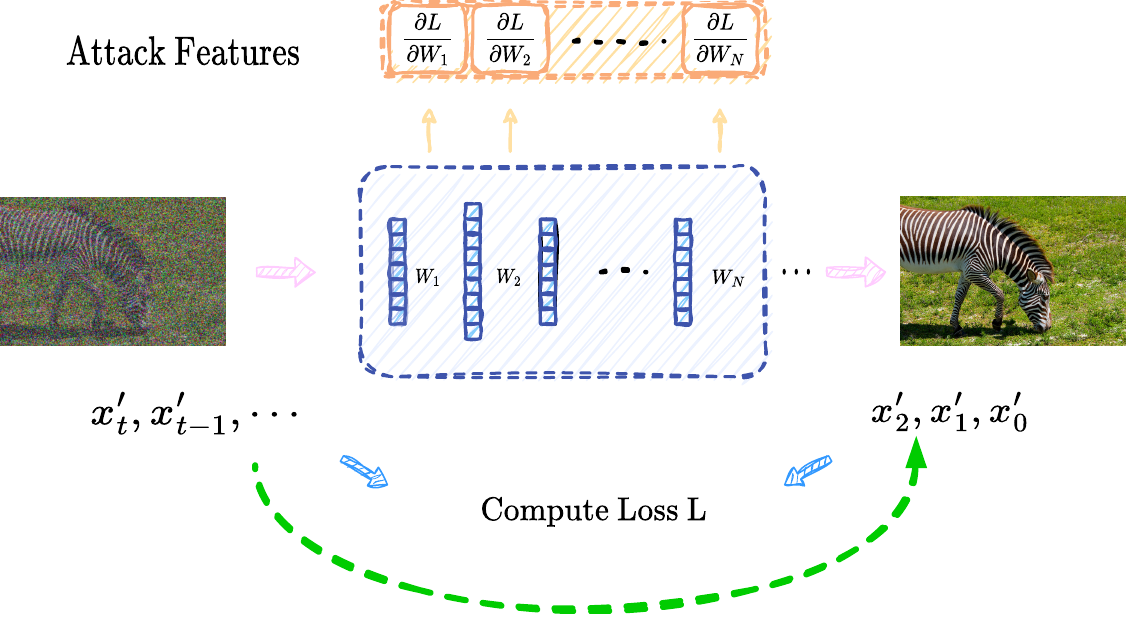}
    \caption{High-level pipeline of our attack: Given the target sample $x_0$, we first add noise based on~\autoref{eq:x_t} and feed it to the target model shaded in blue. At each sample step, we can compute a loss $L$ using~\autoref{eq:diffusion-loss} to derive the gradients. Gradients from all sample steps (with appropriate subsampling and aggregation operations) are used as features to train the attack model for MIA.}
    \label{fig:cal_gradient}
\end{figure}

\autoref{fig:cal_gradient} shows the general idea of our attack. 
It is important to note that, owing to the specific architecture of the diffusion model, a single query point can yield multiple loss values originating from different timesteps. Subsequently, based on the loss $L$, we can derive the gradients using the standard back-propagation technique and use the gradients as features to train a machine-learning model to execute MIA.

{In the diffusion model, the training loss function is defined as~\autoref{eq:diffusion-loss}. For each sample, noise is added using~\autoref{eq:x_t}, generating a noised sample $x_t$. The trained U-Net modules then predict the noise $\epsilon_t$ that needs to be denoised at timestep $t$, based on $x_t$ and $t$. The existing methods~\cite{carlini2023extracting,hu2023membership,matsumoto2023membership} assume that the loss value of a member sample is typically smaller than that of a non-member sample, which indicates intuitively
}
{
\[x \in \mathcal{D}_m \mbox{\; if and only if \;} \lVert \epsilon_t - \epsilon_{\theta}(x_t, t) \rVert^2_2 < \tau\]
}
\noindent where $\epsilon_{\theta}(x_t, t)$ represents the predicted noise at $t$-th step and $\epsilon_t$ is the ground true noise sample.
However, we have observed that this approach can lead to misjudgments. For example, inherently complex member samples might exhibit higher loss values compared to simpler non-member samples, a phenomenon also observed in GAN-Leaks~\cite{chen2020gan}. This indicates that relying solely on loss as the attack feature may introduce some degree of bias. Carlini et al.~\cite{carlini2023extracting} also found that using loss values as the sole criterion for determining membership is inadequate.

{In our work, we propose using gradient values as attack features to better capture the model's reaction to samples. Unlike loss values, which are scalars and provide limited information, gradient data offer a more comprehensive view. Additionally, even when two samples have identical loss values, their corresponding gradients can differ, as gradients depend on the specific inputs within the computational graph. For instance, the diffusion model $\epsilon_{\theta}$ (with parameter $\theta$) calculates gradients for a query sample $x$ at $t$-th step; the gradients can be expressed as:}
{\begin{align}
\nabla_{\theta} L_t(\theta,x) = \nabla_\theta \|\epsilon_t - \epsilon_\theta(x_t, t)\|^2  \label{eq:gra_loss}
\end{align}}
{\hspace{1em}According to the definition of the Euclidean norm squared, we can expand the squared term in~\autoref{eq:gra_loss}:}
{\begin{align*}
\left\lVert \epsilon_t - \epsilon_\theta(x_t, t) \right\rVert^2 
&= \left( \epsilon_t - \epsilon_\theta(x_t, t) \right)^\top \left( \epsilon_t - \epsilon_\theta(x_t, t) \right) \\
&= \lVert \epsilon_t \rVert^2 - 2 \epsilon_t^\top \epsilon_\theta(x_t, t) + \lVert \epsilon_\theta(x_t, t) \rVert^2.
\end{align*}}
{\hspace{1em}Then, we proceed to compute the derivatives of each of the three expanded terms with respect to $\theta$:}

{\begin{align}
\nabla_{\theta} L_t(\theta,x) &= \nabla_{\theta} \left( \left\lVert \epsilon_t \right\rVert^2 - 2\epsilon_t^\top \epsilon_\theta(x_t, t) + \left\lVert \epsilon_\theta(x_t, t) \right\rVert^2 \right) \nonumber \\
&= \nabla_{\theta} \left\lVert \epsilon_t \right\rVert^2 - 2 \nabla_{\theta} \left( \epsilon_t^\top \epsilon_\theta(x_t, t) \right) + \nabla_{\theta} \left\lVert \epsilon_\theta(x_t, t) \right\rVert^2 \nonumber \\
&= 0 - 2\epsilon_t^\top \nabla_{\theta} \epsilon_\theta(x_t, t) + 2 \epsilon_\theta(x_t, t)^\top \nabla_{\theta} \epsilon_\theta(x_t, t) \nonumber \\
&= -2 \left( \epsilon_t - \epsilon_\theta(x_t, t) \right)^\top \nabla_{\theta} \epsilon_\theta(x_t, t) \nonumber \\ 
&= 2 \left( \epsilon_\theta(x_t, t) - \epsilon_t \right)^\top \nabla_{\theta} \epsilon_\theta(x_t, t) \label{eq:method}
\end{align}}

{From~\autoref{eq:method}, we show the gradient depends on both the value of the training loss $\left( \epsilon_\theta(x_t, t) - \epsilon_t \right)$ and the specific query sample being computed $(\nabla_{\theta} \epsilon_\theta(x_t, t))$. For member and non-member samples that produce the same numerical loss value, gradients can still use $\nabla_{\theta} \epsilon_\theta(x_t, t)$ to discriminate them. We also present the experimental evidence to support our finding in~\hyperref[appendix:methodology]{Appendix~\ref*{appendix:methodology}}.

Intuitively, during the training phase, the model fits to member samples. Therefore, when encountering a training sample, the already converged model requires less parameter adjustment compared to a non-member sample, leading to smaller gradients. 
Based on this intuition, we use the model’s gradient values as features for detecting query sample membership, as expressed by:}
{
\[x \in \mathcal{D}_m \mbox{\; if and only if \;} \nabla_{\theta} \lVert \epsilon_t - \epsilon_{\theta}(x_t, t) \rVert^2_2 < \tau\]
}

{The above findings demonstrate that even when the loss values are equal, the gradient information obtained from different samples still varies. We believe that this characteristic of gradient data represents the model's response to the query sample more effectively than the attack features used in existing methods~\cite{carlini2023extracting,matsumoto2023membership,hu2023membership}, thereby enabling more successful attacks.}. However, the key challenge of using gradients for MIA is utilizing gradient information effectively. Considering the substantial number of parameters in the diffusion model (for instance, in our experiments, the Imagen model boasts close to $250$ million 
 trained parameters, while the DDPM model approaches $114$ million), training the attack model by using the gradient of each model parameter for every image is both computationally impractical and prone to overfitting, despite its potential to maximally differentiate between member and non-member samples. Moreover, in diffusion models, the diffusion process typically involves $T$ timesteps (usually set to 1000). For each timestep $t$ in the range $1, \ldots, T$, a separate loss and set of gradients are generated, further increasing the dimensionality of the overall gradients.

\begin{table}[t]
    \centering
    \caption{{Impact of three different timestep-level sampling methods on attack accuracy and their respective time consumption.}}
    \label{tab:case_study}
    \resizebox{0.49\textwidth}{!}{
    \begin{tabular}{cccccc}
    \toprule
         Method & ASR & AUC & TPR@$1\%$FPR & TPR@$0.1\%$FPR & Time (seconds) \\ \midrule
         Effective & $0.947$ & $0.992$ & $0.663$ & $0.311$ & $21587$\\
         Poisson & $0.801$ & $0.882$ & $0.270$ & $0.053$ & $2422$ \\
         Equidistant & $0.932$ & $0.981$ & $0.641$ & $0.304$ & $2398$\\
    \bottomrule
    \end{tabular}} 
\end{table}

\subsection{Gradient Dimensionality Reduction} \label{methodology:reduction and aggregation}
We propose a general attack framework for reducing the dimensionality of the gradients while trying to keep the useful information for differentiating members vs non-members. It is composed of two common techniques: (1) subsampling, which chooses the most informative gradients in a principled way, and (2) aggregation, which combines/compresses those informative gradients data. We name the framework Gradient attack based on Subsampling and Aggregation (GSA).

We then present a three-level taxonomy outlining where these two techniques can be applied: at the timestep level, across different layers within the target model, and within specific gradients of each layer, as detailed below.
\begin{enumerate}[leftmargin=*]
    \item  Timestep Level: As corroborated by prior studies~\cite{carlini2023extracting,duan2023diffusion,hu2023membership,kong2023efficient,matsumoto2023membership}, diffusion models display distinct reactions to member and non-member samples {\it depending on the timestep}. For instance, Carlini et al.~\cite{carlini2023extracting} identified a `Goldilock's zone', which yielded the most effective results in their attack, to be within the range \( t \in [50,300] \). {We believe that the importance of gradient data also varies across different timesteps. Therefore, sampling the timesteps that contain the most useful information will undoubtedly result in more accurate attack outcomes. We refer to the attacks conducted on the most effective gradient data within the `Gold zone' as \textit{effective sampling}. However, implementing \textit{effective sampling} requires detecting the `Gold zone' in the target model each time, and the optimal timesteps for achieving the best attack accuracy may vary across different models. As a result, we propose two alternative sampling methods: \textit{equidistant sampling} and \textit{poisson sampling}. In \textit{equidistant sampling}, the denoising steps are selected at intervals of $T/|K|$ ($K$ refer to the sampled timesteps set) for any given model. In \textit{poisson sampling}, an average rate parameter $\lambda$ ($|K|/T$) is used to randomly generate intervals following an exponential distribution, thereby selecting $|K|$ steps from a total of $T$ steps. We then present a simple case study to test and compare these three different sampling methods.}

    \item Layer-wise Selection and Aggregation: Beyond timesteps, the layers within the model present another pivotal dimension for subsampling and aggregation. Recognizing the nuances captured across layers—from basic patterns in shallower layers to intricate details in deeper ones—it is deemed essential to selectively harness gradients from these layers, especially the informative ones, to optimize the attack model's training.
    
    \item Gradients within Each Layer: Within each layer of a neural network, there is typically no specific ordering of the gradient data. Therefore, it is more reasonable to treat these gradients as a set~\cite{ganju2018property}.  
     
\end{enumerate}

\paragraph{Case Study.} { Since existing attacks~\cite{carlini2023extracting,duan2023diffusion,hu2023membership,kong2023efficient,matsumoto2023membership} heavily focus on timestep-level selection, we designed a case study to better examine how different subsampling methods impact attack performance. We evaluated the attack accuracy using three sampling methods: \textit{effective sampling}, \textit{equidistant sampling}, and \textit{poisson sampling}. For \textit{effective sampling}, it is necessary to first identify the `Gold zone'. To achieve this, we recorded the attack results in every $20$ step across the $T$ denoising steps. The timestep with the best attack performance, along with the $10$ surrounding timesteps, was then selected as the sampling points for effective sampling. For \textit{equidistant sampling}, we set step $1$ as the initial step and then sample timesteps at fixed intervals of $T/|K|$. In contrast, \textit{poisson sampling} uses $|K|/T$ as the parameter $\lambda$ to sample from the $T$ steps.}

{We select $5000$ samples from CIFAR-10 dataset to train DDPM as target model. For each sampling method, we set the number of sampling steps ($|K|$) to $10$. In~\autoref{tab:case_study}, we found that \textit{effective sampling} achieves the highest attack accuracy, while \textit{poisson sampling} has the lowest. This result aligns with our initial assumption that using gradient data sampled from the `Gold zone'—the interval yielding the best attack results on individual timesteps—would lead to optimal performance. In contrast, \textit{poisson sampling}'s randomness may lead to poor attack outcomes if the sampled timesteps cannot effectively discriminate between members and non-members. }

{However, in~\autoref{tab:case_study}, we also present the time consumption for implementing different sampling methods. We found that although \textit{effective sampling} achieves high attack accuracy, it takes nearly $8$ times longer compared to \textit{equidistant} and \textit{poisson sampling}. This is because \textit{effective sampling} requires precomputing the attack performance for numerous timesteps to identify the `Gold zone'. Meanwhile, \textit{equidistant sampling} only slightly reduces the ASR by $0.015$ and the AUC by $0.011$ compared to \textit{effective sampling}, while being more time-efficient. To balance effectiveness and efficiency, we use \textit{equidistant sampling} to derive a subsampled timestep set $K$ from the total diffusion steps $T$.}
Following this, for the timesteps in $K$, we can aggregate the gradients or losses generated at each timestep using statistical methods such as the mean, median, or trimmed mean to produce the final output. If the values being aggregated are the gradients from each timestep, the output can be directly used as the final output. However, if the aggregated values are the losses, the processed loss value needs to be used in backpropagation to extract gradient information for the final output.

\begin{algorithm}[t]
    \caption{$\myattackone$}
    \label{alg:GBA_one}
    \begin{algorithmic}[1]
       \REQUIRE Target model denoted as $\epsilon_{\theta}$ with $N$ layers, a equidistantly selected set of timesteps $K$, and a sample $x$.
    \FOR{$t \in K$}
        \STATE Sample $\epsilon_t$ from Gaussian distribution
        \STATE Compute $x_t$ based on \autoref{eq:x_t}
        \STATE Compute loss $L_t$ from \autoref{eq:diffusion-loss}
    \ENDFOR
        \vspace{0.2cm}
        \STATE $\bar{{L}} \gets \frac{1}{|K|}\sum_{t\in K} {L}_t$ 
        \vspace{0.1cm}
        \STATE $\mathcal{G}\gets \left[ \left\lVert \frac{\partial \bar{{L}}}{\partial W_1} \right\rVert_2^2, \ldots \left\lVert \frac{\partial \bar{{L}}}{\partial W_N} \right\rVert_2^2 \right]$ 
        \vspace{0.1cm}
        \ENSURE $\mathcal{G}$
    \end{algorithmic}
\end{algorithm}

\subsection{Our Instantiations}

We present two exemplary instantiations of the attack within the framework, representing two extreme points in the trade-off space between efficiency and effectiveness. We call them \myattackone and \myattacktwo. \myattackone performs more reduction, gaining efficiency but losing information. \myattacktwo does less reduction, retaining effectiveness but at a cost to efficiency. 
In the \myattackone method, although we equidistantly sample \(|K|\) timesteps from  \(T\), only a single gradient computation is required. 
This outcome is realized by in \myattackone computing the loss, $L_t$, for each timestep present in $K$. Subsequently, we take the mean of these individual losses, represented as $\bar{L}$, to perform backpropagation. This process eventually yields a solitary gradient vector. On the other hand, \myattacktwo entails performing backpropagation and computing gradients for each timestep in $K$, and then using the mean of all gradient vectors, denoted as $\mathcal{G}$, as the final output.

Note that we only slightly optimize our two instantiations in this paper because they are already very effective. We leave more detailed investigations of the design space and more effective proposals as future work.

Based on our detailed analysis of existing white-box attacks~\cite{carlini2023extracting,hu2023membership,matsumoto2023membership}, we first find that the optimal timesteps for mounting the most effective attacks vary depending on the specific dataset and diffusion model in question. 

Consequently, we adopt the \textit{equidistant sampling} strategy to select sample timesteps from the range $[1, T]$, denoted by a set $K$. This approach is designed to encompass timesteps that can distinctly differentiate between member and non-member samples, avoiding an exclusive focus on timesteps that are either too early or too late.

After getting loss from each selected step, we use backpropagation to compute the gradients for the model. Given the diverse nature of gradients within a layer, we aggregate the model's gradient information on a per-layer basis. That is, once the gradient information for a layer's parameters is obtained, the $\ell_2$-norm of these gradients is used as the representation for that layer's gradient information. This approach offers a dual advantage: it substantially reduces computational overhead while also holistically encapsulating that layer's gradient information.

This forms the basis of $\myattacktwo$ (given in~\autoref{alg:GBA_two}): for each timestep $t$ in the set $K$, we calculate the per-layer gradient using the $\ell_2$-norm, and then find their average.

However, this approach can still incur substantial computational costs when applied to large diffusion models and datasets — taking nearly 6 hours to execute on Imagen. To address this inefficiency, we preprocess the loss values from multiple timesteps before doing gradient computation. In light of this challenge, we introduce $\myattackone$ (outlined in~\autoref{alg:GBA_one}), which reduces the gradient extraction time for the Imagen model to less than $2$ hours, significantly decreasing the computational time required.

\begin{algorithm}[t]
    \caption{$\myattacktwo$}
    \label{alg:GBA_two}
    \begin{algorithmic}[1]
       \REQUIRE Target model denoted as $\epsilon_{\theta}$ with $N$ layers, a equidistantly selected set of timesteps $K$, and a sample $x$.
        \STATE $\mathcal{G}\gets[\;]$
        \FOR{$t \in K$}
        \STATE Sample $\epsilon_t$ from Gaussian distribution
        \STATE Compute $x_t$ based on \autoref{eq:x_t}
        \STATE Compute loss $L_t$ from \autoref{eq:diffusion-loss}
        \STATE ${\mathcal{G}_t} = \left[ \left\lVert \frac{\partial {{L_t}}}{\partial W_1} \right\rVert_2^2, \ldots \left\lVert \frac{\partial {{L_t}}}{\partial W_N} \right\rVert_2^2 \right] $
        \ENDFOR
        \vspace{0.1cm}
        \STATE $\mathcal{G}\gets \frac{1}{|K|}\sum_{t\in K} \mathcal{G}_t $
        \ENSURE $\mathcal{G}$ 
        \vspace{0.1cm}
    \end{algorithmic}
\end{algorithm}

\section{Experimental Setup}\label{experimental_setup}


\subsection{Datasets}
\smallskip\noindent

\begin{figure*}[t]
    \centering
    \includegraphics[width=\textwidth]{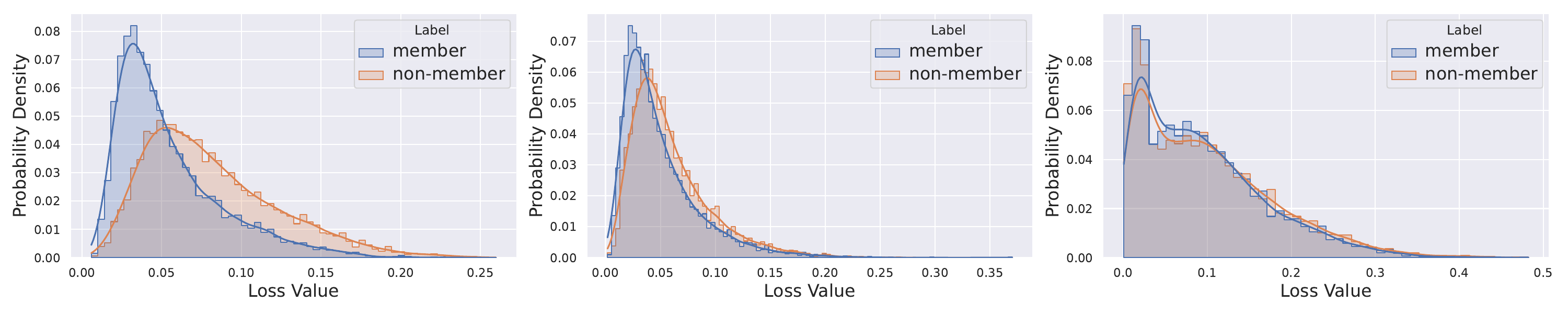}
    \caption{Loss distribution for member vs. non-member samples across CIFAR-10, ImageNet, and MS COCO (from left to right), used by existing work~\cite{hu2023membership,matsumoto2023membership,carlini2023extracting}. Models use default settings from~\autoref{tab:expermients_setup}.}
    \label{dis}
\end{figure*}


We use CIFAR-10, ImageNet, and MS COCO datasets. The use of CIFAR-10 allows for an easier comparison of attack results as it has been frequently employed in previous work~\cite{carlini2023extracting,duan2023diffusion,kong2023efficient,matsumoto2023membership}. Both ImageNet and MS COCO serve as significant target datasets in the domain of image generation, with MS COCO used for training in various tasks, such as VQ-diffusion~\cite{gu2022vector}, Parti Finetuned~\cite{yu2022scaling}, 	
U-ViT-S/2~\cite{bao2023worth}, and Imagen~\cite{saharia2022photorealistic}.

\paragraph{ImageNet} dataset is a large-scale and diverse collection of images designed for image classification and object recognition tasks in the fields of machine learning and computer vision. When conducting experiments with the ImageNet dataset, researchers typically utilize a specific subset consisting of $1.2$ million images for training and 50,000 images for validation, while an additional $100,000$ images are reserved for testing. Considering the constraints on training resources and to ensure diversity in the training images, we opt to utilize the ImageNet test set as the training set for training the models in our work.

\paragraph{CIFAR-10} dataset comprises $10$ categories of $32\times32$ color images, with each category containing $6,000$ images. These categories include airplanes, automobiles, birds, cats, deer, dogs, frogs, horses, ships, and trucks. In total, the dataset consists of $60,000$ images, of which $50,000$ are designated for training and $10,000$ for testing. The CIFAR-10 dataset is commonly employed as a benchmark for image classification and object recognition tasks in the fields of machine learning and computer vision. 

\paragraph{MS COCO} dataset contains over $200,000$ labeled high-resolution images collected from the internet, with a total of 1.5 million object instances and $80$ different object categories. The categories cover a wide range of common objects, including people, animals, vehicles, and household items, among others. The MS COCO dataset is noteworthy for its diversity and the complexity of its images and annotations. Images in the MS COCO dataset depict a wide variety of scenes and object layouts. In this experiment, we utilize all images from the MS COCO training set for model training. The first caption from the five associated with each image is selected as the corresponding textual description.

\subsection{Training Setup}
\begin{table}[h]
    \centering
    \caption{Default parameters used for the experiments.}
    \label{tab:expermients_setup}
    \resizebox{0.47\textwidth}{!}{
    \begin{tabular}{cccc}
        \toprule[1.3pt]
        \multirow{2}{*}{\begin{tabular}[c]{@{}c@{}} Parameters \end{tabular}} & 
        \multirow{2}{*}{\begin{tabular}[c]{@{}c@{}} Unconditional \\ Diffusion \end{tabular}} &
        \multirow{2}{*}{\begin{tabular}[c]{@{}c@{}} Unconditional \\ Diffusion \end{tabular}} &
        \multirow{2}{*}{\begin{tabular}[c]{@{}c@{}} Imagen \end{tabular}}  \\ \\ 
        \midrule  
        \begin{tabular}[c]{@{}c@{}}Channels \end{tabular} & 
        \begin{tabular}[c]{@{}c@{}}$128$ \end{tabular} &
        \begin{tabular}[c]{@{}c@{}}$128$ \end{tabular} &
        \begin{tabular}[c]{@{}c@{}}$128$ \end{tabular} 
        \\ 
        \begin{tabular}[c]{@{}c@{}} Diffusion steps \end{tabular} &
        \begin{tabular}[c]{@{}c@{}}$1000$ \end{tabular} & 
        \begin{tabular}[c]{@{}c@{}}$1000$ \end{tabular} &
        \begin{tabular}[c]{@{}c@{}}$1000$ \end{tabular} 
        \\ 
        \begin{tabular}[c]{@{}c@{}} Dataset \end{tabular} & 
        \begin{tabular}[c]{@{}c@{}}CIFAR-10 \end{tabular} &
        \begin{tabular}[c]{@{}c@{}}ImageNet \end{tabular} &
        \begin{tabular}[c]{@{}c@{}}MS COCO \end{tabular} 
        \\ 
        \begin{tabular}[c]{@{}c@{}}Training data size \end{tabular} & 
        \begin{tabular}[c]{@{}c@{}}$8000$ \end{tabular} &
        \begin{tabular}[c]{@{}c@{}}$8000$ \end{tabular} &
        \begin{tabular}[c]{@{}c@{}}$30000$ \end{tabular} 
        \\ 
        \begin{tabular}[c]{@{}c@{}}Resolution \end{tabular} &
        \begin{tabular}[c]{@{}c@{}}$32$ \end{tabular} &
        \begin{tabular}[c]{@{}c@{}}$64$ \end{tabular} &
        \begin{tabular}[c]{@{}c@{}}$64$ \end{tabular} 
        \\ 
        \begin{tabular}[c]{@{}c@{}}Learning rate \end{tabular} &
        \begin{tabular}[c]{@{}c@{}}$1e-4$\end{tabular} &
        \begin{tabular}[c]{@{}c@{}}$1e-4$\end{tabular} &
        \begin{tabular}[c]{@{}c@{}}$1e-4$\end{tabular}
        \\ 
        \begin{tabular}[c]{@{}c@{}}Batch size\end{tabular} &
        \begin{tabular}[c]{@{}c@{}}$64$\end{tabular} &
        \begin{tabular}[c]{@{}c@{}}$64$\end{tabular} &
        \begin{tabular}[c]{@{}c@{}}$64$\end{tabular}
        \\  
        \begin{tabular}[c]{@{}c@{}}Noise schedule \end{tabular} & 
        \begin{tabular}[c]{@{}c@{}}linear\end{tabular} &
        \begin{tabular}[c]{@{}c@{}}linear\end{tabular} &
        \begin{tabular}[c]{@{}c@{}}linear, cosine\end{tabular} 
        \\ 
        \begin{tabular}[c]{@{}c@{}}Learning rate schedule\end{tabular} &
        \begin{tabular}[c]{@{}c@{}}cosine\end{tabular} &
        \begin{tabular}[c]{@{}c@{}}cosine\end{tabular} &
        \begin{tabular}[c]{@{}c@{}}cosine\end{tabular} \\ 
         \begin{tabular}[c]{@{}c@{}}Training time\end{tabular} &
         \begin{tabular}[c]{@{}c@{}}$400$ epochs\end{tabular} &
         \begin{tabular}[c]{@{}c@{}}$400$ epochs\end{tabular} & 
         \begin{tabular}[c]{@{}c@{}}$600,000$ steps\end{tabular} \\
        \bottomrule[1.3pt]
    \end{tabular}}
\end{table}
We tabulated the default training parameters for the unconditional diffusion model on CIFAR-10 and ImageNet, and for Imagen on MS COCO, in~\autoref{tab:expermients_setup}. Given that we have employed ASR (Accuracy) as our evaluation metric, we endeavor to maintain a balance between the quantities of the member set and the non-member set to ensure the precision of model validation. The structure for the unconditional diffusion model aligns with those from the diffusers library~\cite{von-platen-etal-2022-diffusers} in Huggingface. Imagen is based on the open-source implementation by Phil Wang et al.\footnote{Code available at \url{https://github.com/lucidrains/imagen-pytorch}}, and we have retained consistency in its configuration. All experiments were conducted using two NVIDIA A100 GPUs.

\subsection{Metrics}
\smallskip\noindent
In the process of comparing experimental results, we employ Attack Success Rate (ASR)~\cite{choquette2021label}, Area Under the ROC Curve (AUC), and True-Positive Rate (TPR) values under fixed low False-Positive Rate (FPR) as evaluation metrics. 

In our experiments, we ensure an equal number of member and non-member image samples. Given the balanced nature of our dataset and the stability of ASR in such contexts, we employ ASR as our primary evaluation metric.

We note that most instances MIAs on diffusion models use the AUC metric for evaluation~\cite{carlini2023extracting,duan2023diffusion,hu2023membership,kong2023efficient,matsumoto2023membership}. Likewise, in assessing the merits of our work in~\autoref{compare_existing}, we will also use AUC as one of our assessment metrics. Additionally, as Carlini et al.~\cite{carlini2022membership} argued that TPR under a low FPR scenario is a key evaluation criterion, we also use TPR at $1\%$ FPR and $0.1\%$ FPR, respectively.


\begin{figure*}[t]
    \centering
    \begin{subfigure}{0.49\textwidth}
        \centering
        \begin{subfigure}{0.31\textwidth}
        \includegraphics[width=1\linewidth]{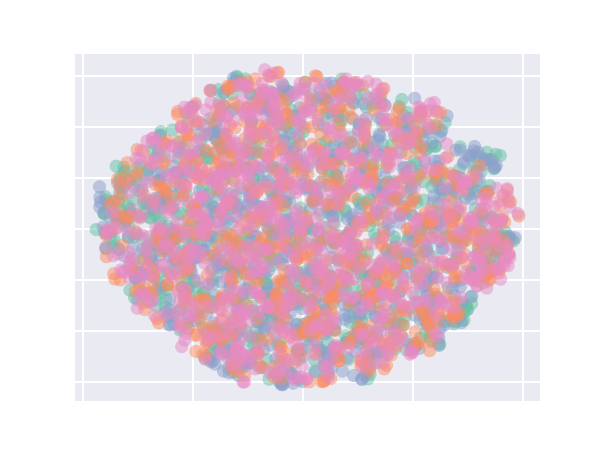}  
        \caption*{epoch : 100}
        \end{subfigure}
        \begin{subfigure}{0.31\textwidth}
            \includegraphics[width=1\linewidth]{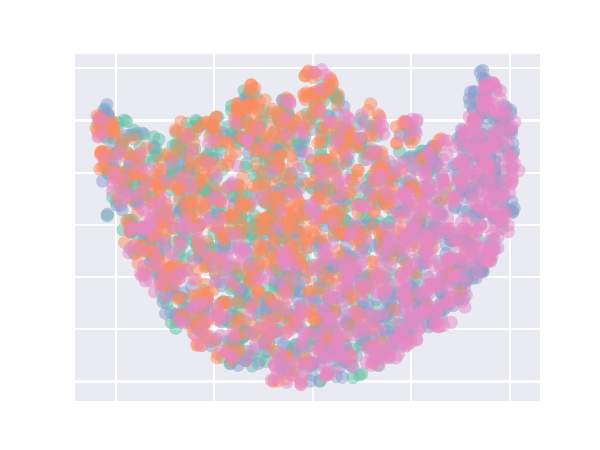}  
        \caption*{epoch : 250}
        \end{subfigure}
        \begin{subfigure}{0.31\textwidth}
            \includegraphics[width=1\linewidth]{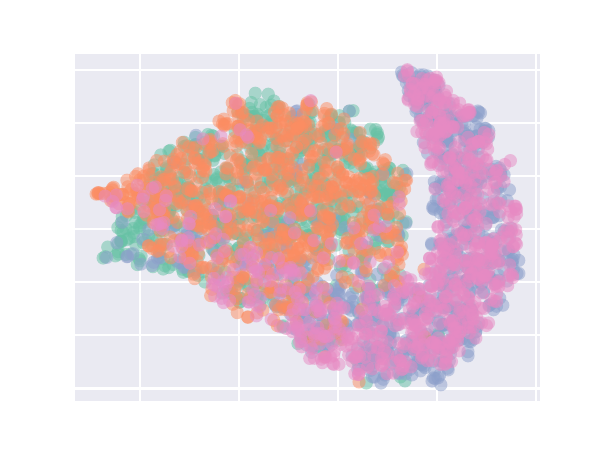}  
        \caption*{epoch : 400}
        \end{subfigure}
        \caption{\myattackone on CIFAR-10}
    \end{subfigure}
    \begin{subfigure}{0.49\textwidth}
        \centering
        \begin{subfigure}{0.31\textwidth}
            \includegraphics[width=1\linewidth]{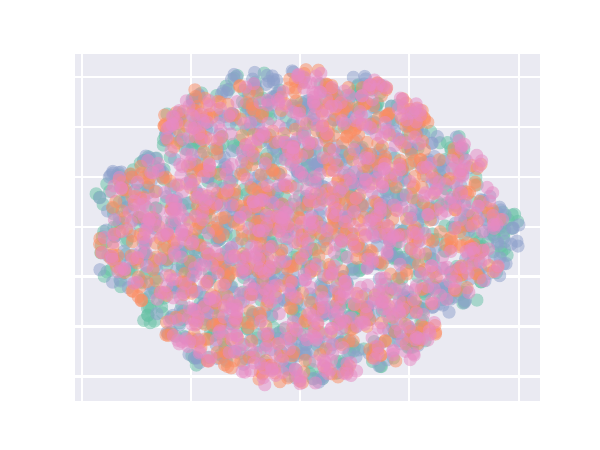}  
        \caption*{epoch : 100}
        \end{subfigure}
        \begin{subfigure}{0.31\textwidth}
            \includegraphics[width=1\linewidth]{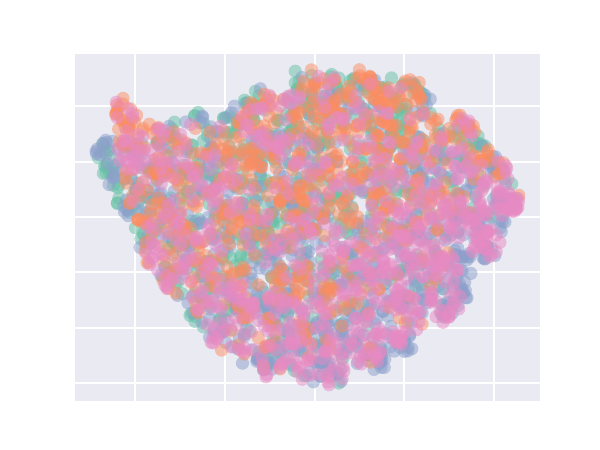}  
        \caption*{epoch : 250}
        \end{subfigure}
        \begin{subfigure}{0.31\textwidth}
            \includegraphics[width=1\linewidth]{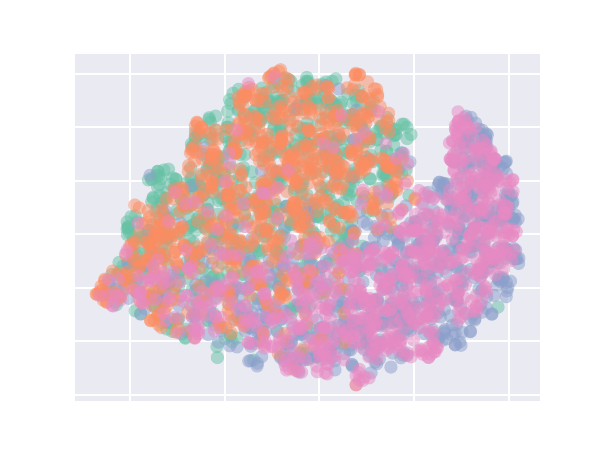}  
        \caption*{epoch : 400}
        \end{subfigure}
        \caption{\myattacktwo on CIFAR-10}
    \end{subfigure}
    \\
    \begin{subfigure}{0.49\textwidth}
        \centering
        \begin{subfigure}{0.31\textwidth}
        \includegraphics[width=1\linewidth]{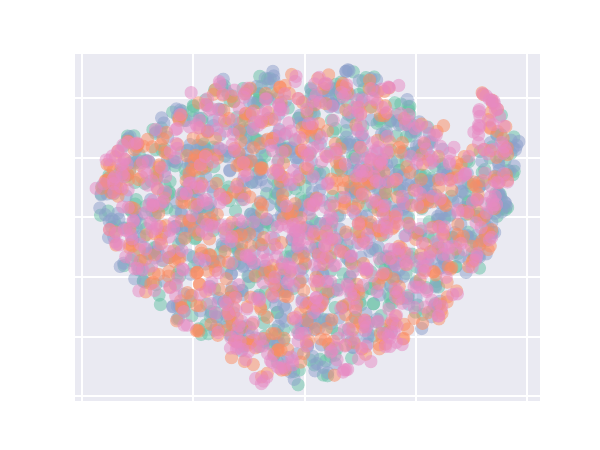}  
        \caption*{epoch: 100}
        \end{subfigure}
        \begin{subfigure}{0.31\textwidth}
            \includegraphics[width=1\linewidth]{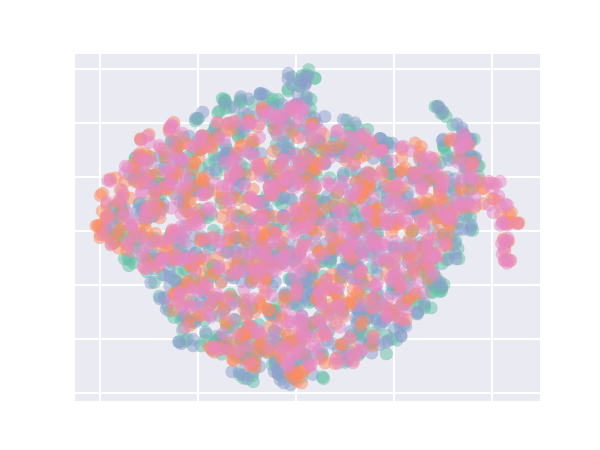}  
        \caption*{epoch: 250}
        \end{subfigure}
        \begin{subfigure}{0.31\textwidth}
            \includegraphics[width=1\linewidth]{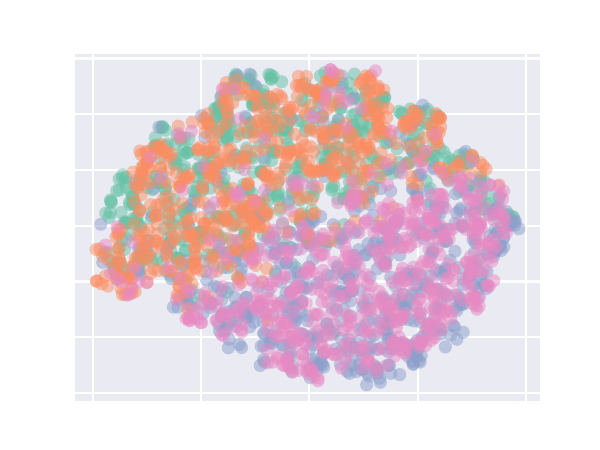}  
        \caption*{epoch: 400}
        \end{subfigure}
        \caption{\myattackone on ImageNet}
    \end{subfigure}
    \begin{subfigure}{0.49\textwidth}
        \centering
        \begin{subfigure}{0.31\textwidth}
            \includegraphics[width=1\linewidth]{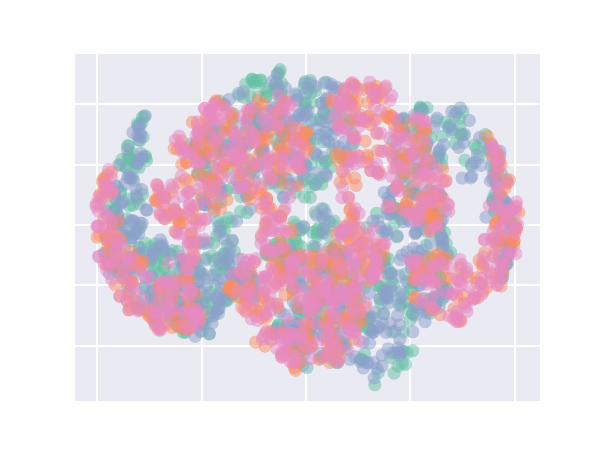}  
        \caption*{epoch: 100}
        \end{subfigure}
        \begin{subfigure}{0.31\textwidth}
            \includegraphics[width=1\linewidth]{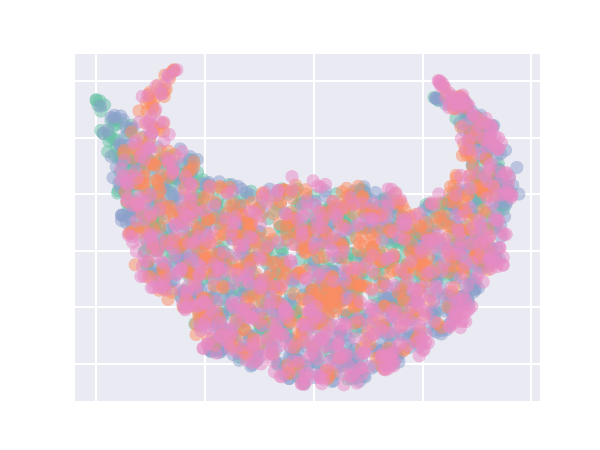}  
        \caption*{epoch: 250}
        \end{subfigure}
        \begin{subfigure}{0.31\textwidth}
            \includegraphics[width=1\linewidth]{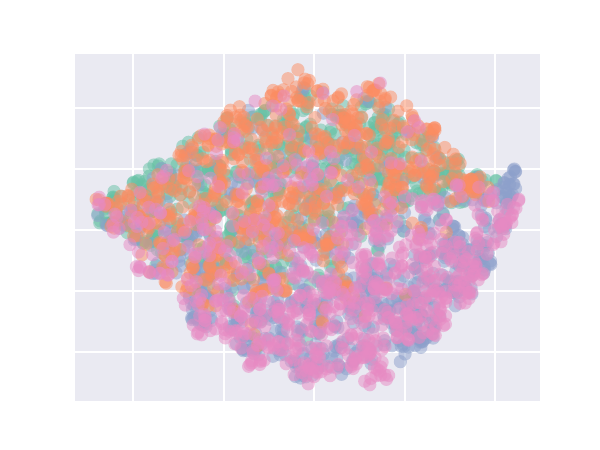}  
        \caption*{epoch: 400}
        \end{subfigure}
        \caption{\myattacktwo on ImageNet}
    \end{subfigure}
    \\
    \begin{subfigure}{0.49\textwidth}
        \centering
        \begin{subfigure}{0.31\textwidth}
        \includegraphics[width=1\linewidth]{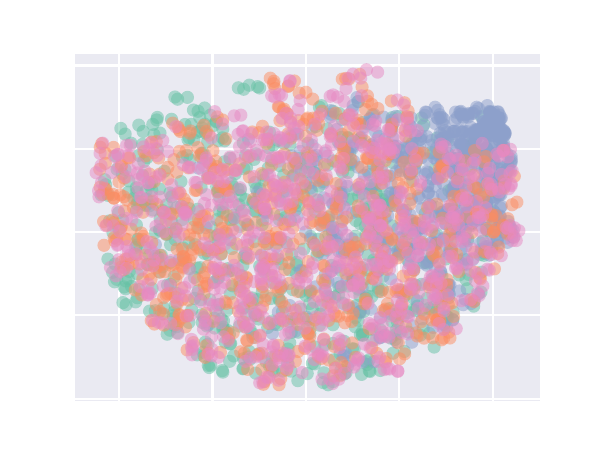}  
        \caption*{step: 200000}
        \end{subfigure}
        \begin{subfigure}{0.31\textwidth}
            \includegraphics[width=1\linewidth]{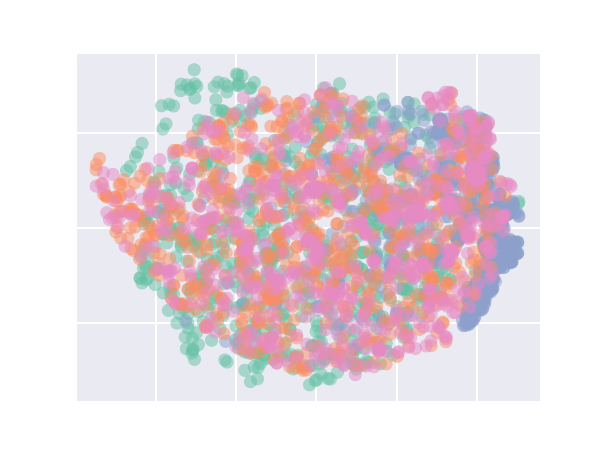}  
        \caption*{step: 400000}
        \end{subfigure}
        \begin{subfigure}{0.31\textwidth}
            \includegraphics[width=1\linewidth]{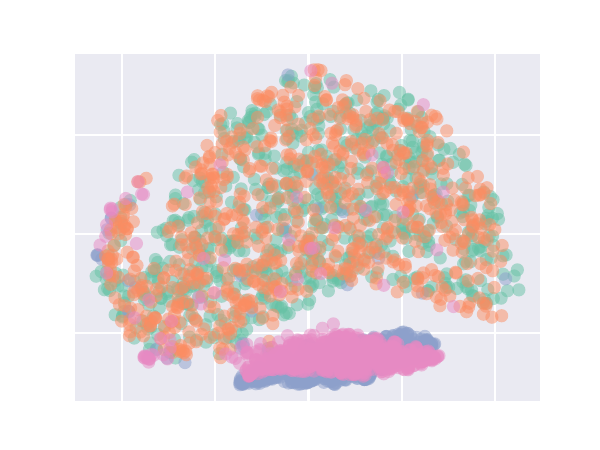}  
        \caption*{step: 600000}
        \end{subfigure}
        \caption{\myattackone on MS COCO}
    \end{subfigure}
    \begin{subfigure}{0.49\textwidth}
        \centering
        \begin{subfigure}{0.31\textwidth}
        \includegraphics[width=1\linewidth]{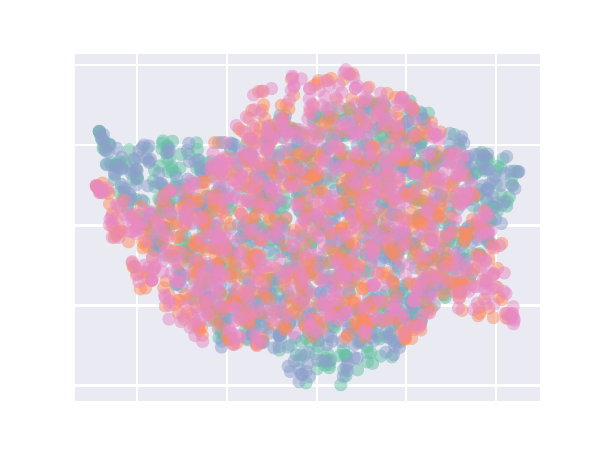}  
        \caption*{step: 200000}
        \end{subfigure}
        \begin{subfigure}{0.31\textwidth}\includegraphics[width=1\linewidth]{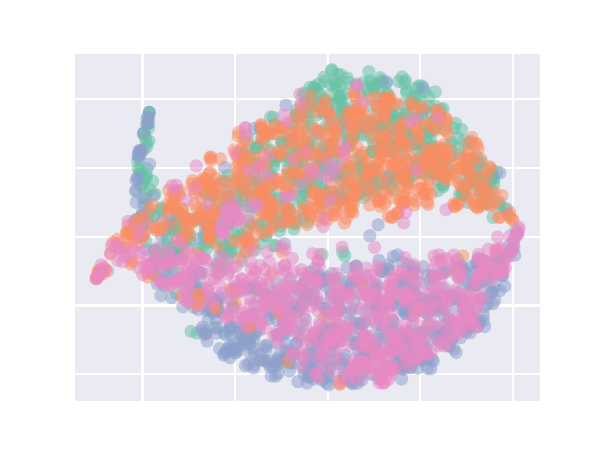}  
        \caption*{step: 400000}
        \end{subfigure}
        \begin{subfigure}{0.31\textwidth}
        \includegraphics[width=1\linewidth]{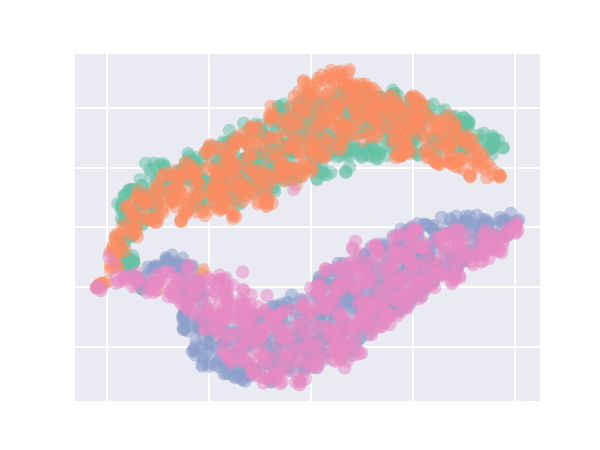}  
        \caption*{step: 600000}
        \end{subfigure}
        \caption{\myattacktwo on MS COCO}
    \end{subfigure}
    \\
    \begin{subfigure}{0.97\textwidth}
    \centering
    \begin{tikzpicture}
        \newlength{\mywidth}
        \setlength{\mywidth}{3.5cm} 
        \node[draw=shadow_member, fill=shadow_member, circle, inner sep=2pt, label={[black,xshift=0.1cm]right:shadow member}] (A) at (0,0) {};
        \node[draw=shadow_non-member, fill=shadow_non-member, circle, inner sep=2pt, label={[black,xshift=0.1cm]right:shadow non-member}, right=\mywidth of A] (B) {};
        \node[draw=target_member, fill=target_member, circle, inner sep=2pt, label={[black,xshift=0.1cm]right:target member}, right=\mywidth of B] (C) {};
        \node[draw=target_non-member, fill=target_non-member, circle, inner sep=2pt, label={[black,xshift=0.1cm]right:target non-member}, right=\mywidth of C] (D) {};
    \end{tikzpicture}
\end{subfigure}
    \caption{The left and right columns display the visualization of high-dimensional gradient information using t-SNE after \myattackone and \myattacktwo have respectively executed attacks on the three datasets (using the output from the last layer of our attack model). For all six attacks, it is observed that member and non-member samples are distinctly differentiated when reaching the training steps defined by the default settings (as referenced in~\autoref{tab:expermients_setup}).}
    \label{gradient_dis}
    
\end{figure*}

\section{Evaluation Results}\label{evaluation_results}
\subsection{Comparison with Existing Methods}\label{compare_existing}
\smallskip\noindent


\begin{table}[h]
    \centering
    \caption{Existing white-box attacks on the CIFAR-10 dataset are benchmarked using four distinct metrics. LiRA\textsuperscript{*}, LSA\textsuperscript{*}, \myattackone, and \myattacktwo are all obtained under the same conditions.}
    \label{table:compare_loss}
     \resizebox{0.47\textwidth}{!}{
     \large
    \begin{tabular}{ccccc}
    \toprule
    \multirow{2}{*}{\begin{tabular}[c]{@{}c@{}} \textbf{Attack} \\ \textbf{method}\end{tabular}} & 
    \multicolumn{4}{c}{\textbf{CIFAR-10}} 
    \\ \cmidrule{2-5} 
    &
    \begin{tabular}[c]{@{}c@{}}\textbf{ASR}$^{\uparrow}$\end{tabular} & 
    \begin{tabular}[c]{@{}c@{}}\textbf{AUC}$^{\uparrow}$\end{tabular} & 
    \begin{tabular}[c]{@{}c@{}}\textbf{TPR@1\%FPR(\%)}$^{\uparrow}$\end{tabular} & 
    \begin{tabular}[c]{@{}c@{}}\textbf{TPR@0.1\%FPR(\%)}$^{\uparrow}$\end{tabular} 
    \\ \cmidrule{1-1} \cmidrule(lr){2-2} \cmidrule(lr){3-3} \cmidrule(lr){4-4} \cmidrule(lr){5-5} 
    \begin{tabular}[c]{@{}c@{}}Baseline\end{tabular} & 
    \begin{tabular}[c]{@{}c@{}}$0.736$\end{tabular} & 
    \begin{tabular}[c]{@{}c@{}}$0.801$\end{tabular} & 
    \begin{tabular}[c]{@{}c@{}}$5.65$\end{tabular} & 
    \begin{tabular}[c]{@{}c@{}}$-$\end{tabular}
    \\ 
    \begin{tabular}[c]{@{}c@{}}LiRA\end{tabular} & 
    \begin{tabular}[c]{@{}c@{}}$-$\end{tabular} & 
    \begin{tabular}[c]{@{}c@{}}$0.982$\end{tabular} & 
    \begin{tabular}[c]{@{}c@{}}$5(5M)\;99(102M)$\end{tabular} & 
    \begin{tabular}[c]{@{}c@{}}$7.1$\end{tabular}
    \\  
    \begin{tabular}[c]{@{}c@{}}Strong LiRA\end{tabular}& 
    \begin{tabular}[c]{@{}c@{}}$-$\end{tabular} & 
    \begin{tabular}[c]{@{}c@{}}$0.997$\end{tabular} & 
    \begin{tabular}[c]{@{}c@{}}$-$\end{tabular}& 
    \begin{tabular}[c]{@{}c@{}}$29.4$\end{tabular}
    \\ 
    \begin{tabular}[c]{@{}c@{}}LiRA\textsuperscript{*}\end{tabular} & 
    \begin{tabular}[c]{@{}c@{}}$0.626$\end{tabular} & 
    \begin{tabular}[c]{@{}c@{}}$0.71$\end{tabular} & 
    \begin{tabular}[c]{@{}c@{}}$1.45$\end{tabular} & 
    \begin{tabular}[c]{@{}c@{}}$0.25$\end{tabular} 
    \\ 
    \begin{tabular}[c]{@{}c@{}}LSA\textsuperscript{*}\end{tabular} & 
    \begin{tabular}[c]{@{}c@{}}$0.83$\end{tabular} & 
    \begin{tabular}[c]{@{}c@{}}$0.909$\end{tabular}& 
    \begin{tabular}[c]{@{}c@{}}$13.77$\end{tabular} & 
    \begin{tabular}[c]{@{}c@{}}$0.925$\end{tabular}
    \\ \midrule
    \rowcolor{gray!25}$\myattackone$& 
    \begin{tabular}[c]{@{}c@{}}$\textbf{0.993}$\end{tabular} & 
    \begin{tabular}[c]{@{}c@{}}$\textbf{0.999}$\end{tabular} & 
    \begin{tabular}[c]{@{}c@{}}$\textbf{99.7}$\end{tabular} & 
    \begin{tabular}[c]{@{}c@{}}$\textbf{82.9}$\end{tabular} 
    \\ 
    \rowcolor{gray!25}$\myattacktwo$ & 
    \begin{tabular}[c]{@{}c@{}}$\textbf{0.988}$\end{tabular} & 
    \begin{tabular}[c]{@{}c@{}}$\textbf{0.999}$\end{tabular} & 
    \begin{tabular}[c]{@{}c@{}}$\textbf{97.88}$\end{tabular} & 
    \begin{tabular}[c]{@{}c@{}}$\textbf{58.57}$\end{tabular} 
    \\ \bottomrule
    \end{tabular}
    }
    
\end{table}


We benchmark $\myattackone$ and $\myattacktwo$ against existing methodologies, maintaining all other model parameters consistent. Contrasting traditional loss-based white-box attacks such as LiRA~\cite{carlini2023extracting} and others techniques~\cite{hu2023membership,matsumoto2023membership}, we provide a thorough evaluation highlighting the superior efficacy of $\myattackone$ and $\myattacktwo$. The baseline approach~\cite{yeom2018privacy,hu2023membership,matsumoto2023membership} depicted in~\autoref{table:compare_loss} is the most intuitive, which predicts the sample as a non-member if the loss exceeds a certain value and vice versa~\cite{matsumoto2023membership}. This also represents the most traditional judgment method in MIA, utilized here as the baseline. 
\subsubsection{Feature Informative}
LSA\textsuperscript{*} refers to the results of training the attack model using the loss under the same training conditions and sampling frequency as $\myattackone$ and $\myattacktwo$. The sole distinction between LSA\textsuperscript{*} and GSA lies in their features: while LSA\textsuperscript{*} utilizes loss as its attack feature, GSA employs the gradient. Comparative results between them substantiate that the gradient information of the diffusion model is more aptly suited as attack features.

It is apparent from~\autoref{table:compare_loss} that both $\myattackone$ and $\myattacktwo$ exceed other techniques in terms of all evaluation metrics. Under the AUC criterion, LiRA~\cite{carlini2023extracting} also attains a high attack accuracy, attributed to excessive training steps and many shadow models. However, when ensuring an equivalent quantity of shadow models and training epochs for the LiRA$\textsuperscript{*}$ based on the LiRA framework, its ASR, TPR, and AUC scores are significantly lower compared to \myattackone and \myattacktwo. In the original paper, the LiRA framework achieves TPRs of $5\%$ after training for $200$ epochs, with the FPRs fixed at $1\%$. Remarkably, after training for $4080$ epochs, the TPR increases to $99\%$. In contrast, for $\myattackone$ and $\myattacktwo$, TPRs of $99.7\%$ and $78.75\%$ are respectively achieved after only $400$ epochs, underscoring a more efficient attack strategy. This essentially corroborates our core proposition that gradient information of the model exhibits a more pronounced response to member set samples than loss. 


\subsubsection{Timestep Selection}

Moreover, the `time zone' demonstrating discernible differences in the loss distribution between members and non-members vary across different models and datasets~\cite{carlini2023extracting,hu2023membership,matsumoto2023membership,duan2023diffusion}. Consequently, to achieve a more potent attack, it becomes imperative to extract the loss and establish thresholds or distributions for each timestep using shadow models, aiming to pinpoint the most efficacious `time zone'. In contrast, both \myattackone and \myattacktwo execute attacks by solely harnessing the gradient information derived from \textit{equidistant sampling} timesteps across the $T$ diffusion steps, achieving similar attack accuracy in just \textbf{one-thirtieth} of the time. Given a consistent dataset size and model architecture, extracting loss across $T$ steps takes 36 hours. In contrast, \myattackone and \myattacktwo achieve the \textbf{same accuracy level} in less than $1$ hour by extracting gradients from $10$ equidistant sampling timesteps.

\begin{figure*}[t]
    \begin{subfigure}{0.66\textwidth}
    \begin{subfigure}{0.49\textwidth}
        \includegraphics[width = \textwidth]{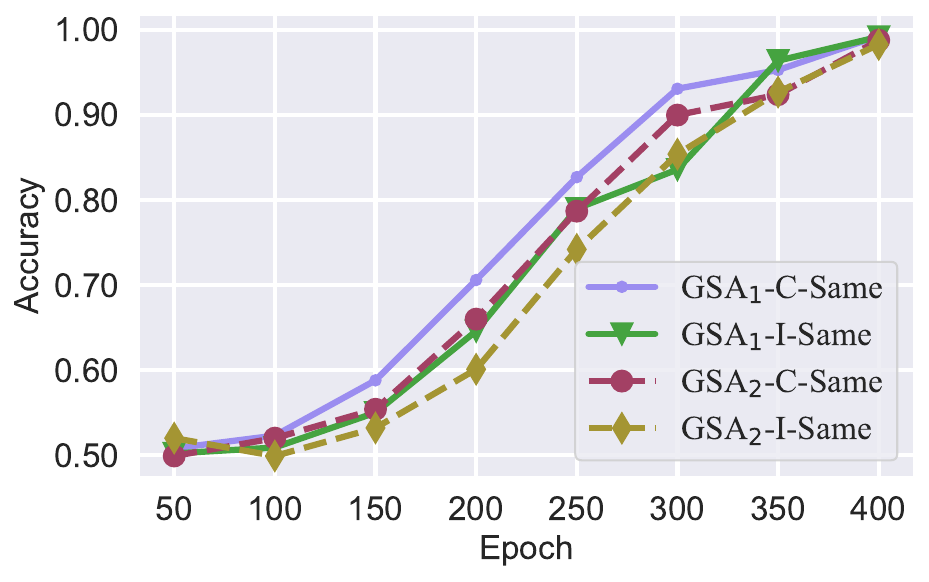}
    \end{subfigure}
    \begin{subfigure}
        {0.49\textwidth}
        \includegraphics[width = \textwidth]{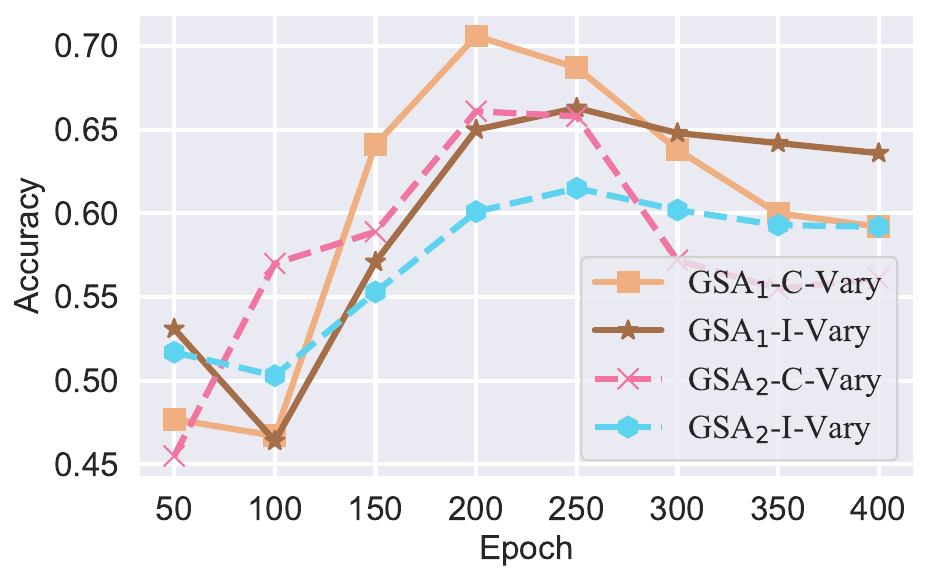}
    \end{subfigure}
    \caption{Impact of training epoch}
    \label{fig:4_a}
    \end{subfigure}
    \begin{subfigure}{0.33\textwidth}
        \includegraphics[width=\textwidth]{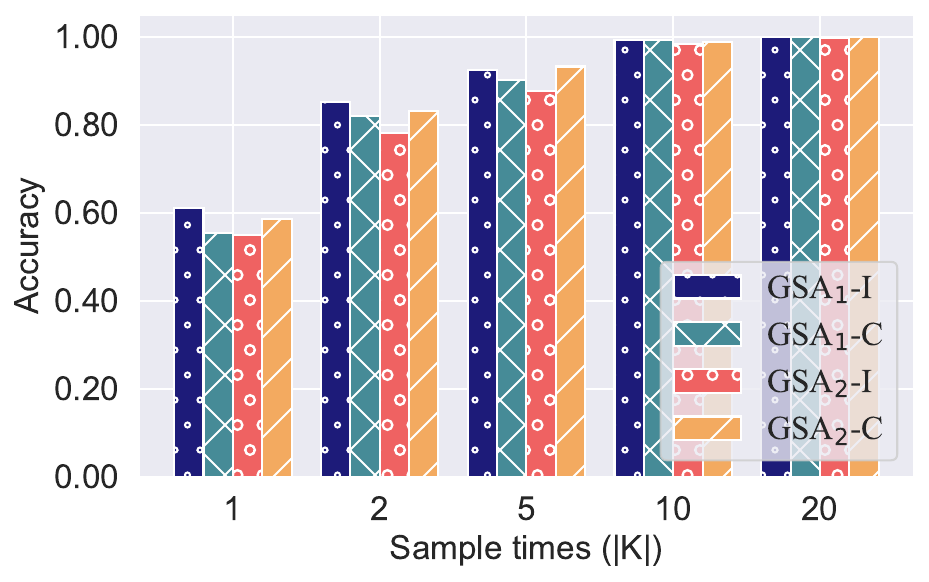}
        \caption{Impact of |K|}
        \label{fig:4_b}
    \end{subfigure}
    \caption{``-I-'' and ``-C-'' denote experiments with ImageNet and CIFAR-10 datasets. Panel (a) (left) reveals that attacks are more effective when shadow and target models closely fit the training data; (right) however, increased fitting disparities between them weaken the attack. Panel (b) shows that greater sampling frequency boosts the attack's effectiveness, possibly due to acquiring finer data and getting more informative timestep.}
    \label{fig:4}
\end{figure*}

To further demonstrate that the optimal timestep for distinguishing between member and non-member samples using loss varies across different datasets and models. We plot the loss distribution for three distinct datasets used in our experiment: CIFAR-10, ImageNet, and MS COCO. Following the methodology of LiRA in attacking diffusion models~\cite{carlini2023extracting}, we identified the optimal timestep for each of the three distinct datasets that best distinguishes member from non-member samples. For this, we equidistantly sampled $10$ timesteps from shadow models (the training times of these shadow models align with those presented in~\autoref{tab:expermients_setup}). However, we observed that the identified timesteps across the three datasets were not consistent. Upon visualizing the loss distribution at these specific timesteps in~\autoref{dis}, we found that even at these optimal points, the loss distribution did not effectively differentiate between member and non-member samples. DDPM trained on the CIFAR-10 dataset clearly differentiates between member and non-member loss distributions. However, such a difference is not pronounced for models trained on ImageNet and MS COCO datasets. For models to execute attacks on the ImageNet and MS COCO datasets, it is essential to compute the loss distribution across a broader range of timesteps and increase their training time. 

Using the same model parameters and sampling frequency as in \autoref{dis}, we tried attacks with \myattackone and \myattacktwo. The attack features were derived from the gradients of timesteps sampled from $T$ using the same sampling frequency as previously employed. We visualized this high-dimensional gradient information using t-SNE~\cite{Maaten2008VisualizingDU} in~\autoref{gradient_dis}. It can be observed quite intuitively, that across all datasets, both \myattackone and \myattacktwo can effortlessly differentiate between target member and target non-member data using the features derived from the gradients of shadow models.


\subsection{Attacking Unconditional Diffusion Model}\label{unconditional}
\smallskip\noindent

In this section, we trained six shadow models to facilitate the attack. We focus on unconditional diffusion models and test on CIFAR-10 and ImageNet datasets. 

\paragraph{Training on Different Epochs.}
Our first goal is to understand how varying training epochs for target and shadow models influence our attacks. We considered two possible scenarios. 
\begin{itemize}[leftmargin=*]
    \item In the first scenario, the attacker knows the target model's training epochs and matches the shadow model's training accordingly.
    \item In the second scenario, the attacker is unaware of the target model's training details and varies only the shadow model's training epochs for experimentation.
\end{itemize}


In~\autoref{fig:4_a}, we present the experimental results under the first scenario. These findings indicate that as the training epochs for both the target and shadow models increase, the attack success rate for $\myattackone$ and $\myattacktwo$ consistently improves. In this context, the suffixes ``-I-'' and ``-C-'' refer to experiments on ImageNet and CIFAR-10, respectively. We postulate that with an increasing number of epochs, the model tends to fit the training data more closely after convergence. This amplifies the gradient discrepancy between member and non-member samples, subsequently bolstering the efficacy of the attack.

In the second scenario setting, when the training epochs of the target model are fixed at $200$ epochs, the attack accuracy is optimal when the shadow model's training epochs closely match those of the target model. Furthermore, observations from~\autoref{fig:4_a} suggest that the overall efficacy of membership inference attacks is closely tied to the consistency in the degree of fit between the shadow models and their training data as compared to that of the target model with its training data. When shadow models exceed the target model in data fitting, it does not invariably lead to an improved attack performance. Contrarily, the attack's success rate might diminish due to disparities in their fitting levels.

Then, our experiments explore the influence of the degree of overfitting in both shadow and target models on attack accuracy. Moreover, we examine the impact of discrepancies in data-fitting levels between the target and shadow models on the performance of the attack.



\paragraph{Sampling Frequency Variation Analysis.}
In both \myattackone and \myattacktwo, the term `sample times' ($\lvert K \rvert$) refers to the number of elements in the set $K$, derived through the \textit{equidistant sampling} of timesteps from $T$. \myattackone and \myattacktwo employ statistical methods on distinct pieces of information; the former determines the mean loss over the $|K|$ timesteps, while the latter computes the average gradient value. Our initial hypothesis was that an increased number of sampling instances, providing the attack model with more information and potentially capturing distinct timesteps that clearly differentiate between member and non-member samples, would lead to improved attack accuracy.


\begin{figure*}
    \begin{subfigure}
        {0.33\textwidth}
        \includegraphics[width = \textwidth]{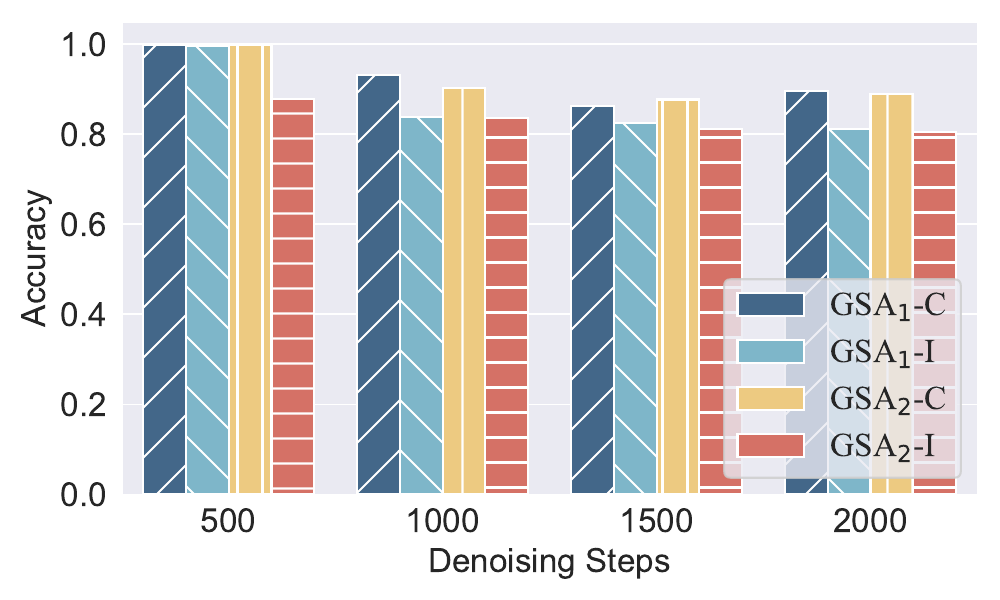}
        \caption{Impact of diffusion steps}
        \label{fig:5_a}
    \end{subfigure}
    \begin{subfigure}
        {0.33\textwidth}
        \includegraphics[width = \textwidth]{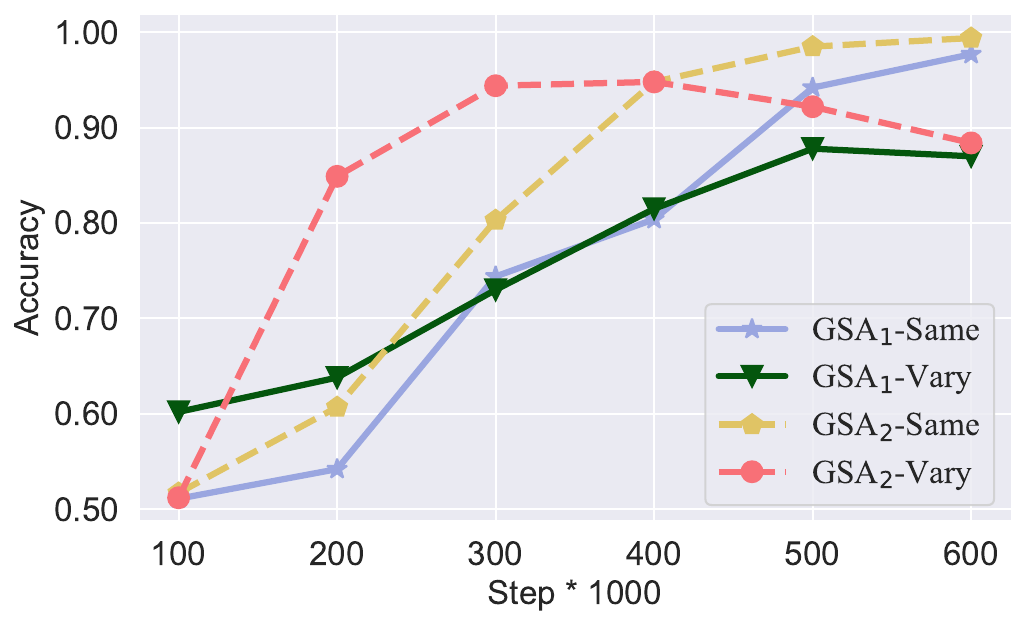}
        \caption{Impact of training epoch}
        \label{fig:5_b}
    \end{subfigure}
    \begin{subfigure}
        {0.33\textwidth}
        \includegraphics[width = \textwidth]{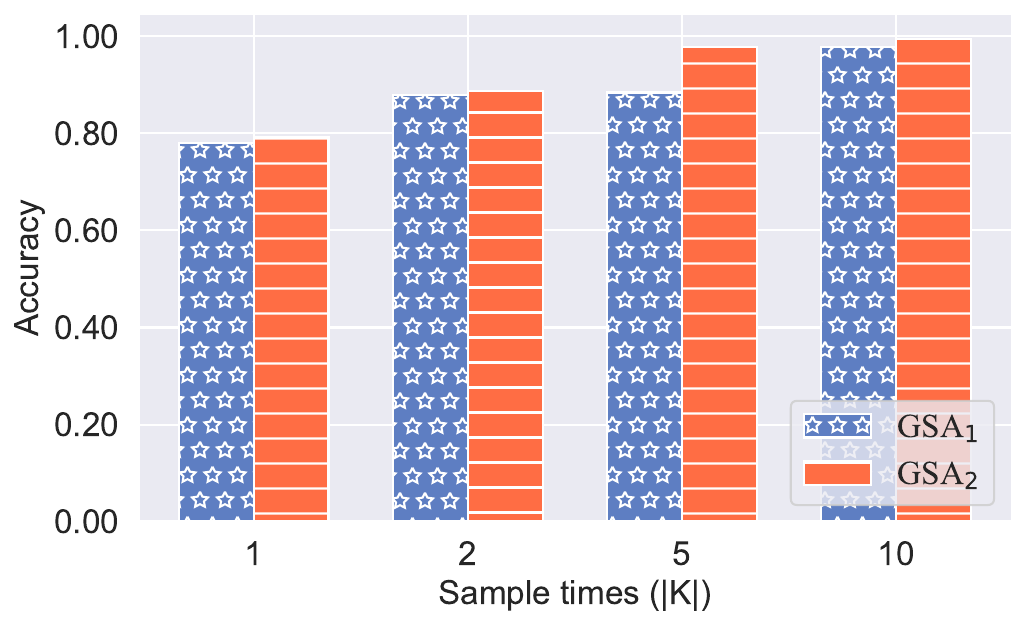}
        \caption{Impact of |K|}
        \label{fig:5_c}
    \end{subfigure}
    \caption{{Notations ``-I-'' and ``-C-'' are consistent with those in~\autoref{fig:4_a}. Panel (a) suggests that increasing the number of diffusion steps, which decelerates convergence, results in a reduced attack success rate.} Panel (b) reinforces findings from~\autoref{fig:4_a}: enhanced data-fitting by both the shadow and target models boosts the attack's efficacy. However, when there are disparities in the data fitting, the efficacy diminishes. Panel (c) shows that augmenting the sampling steps for Imagen—thus acquiring more information—significantly improves the attack's success rate.}
    \label{fig:5}
\end{figure*}

~\autoref{fig:4_b} confirms our initial hypothesis that collecting more gradient information from a single sample enhances the attack's success rate. In all attacks, we maintained a constant setting of 1000 diffusion steps and conducted equidistant sampling across these steps. Our focus was on understanding how varying the sampling frequencies during the evaluation process of a single sample affects the attack's accuracy.

From our experimental data, we observed that the attack's success rate was lowest when gradient information was collected only once per sample. This limited data collection blurred the distinction between member and non-member set samples. Notably, the precision saw a significant boost when the collection frequency increased. However, after reaching a threshold of ten collections per sample, further increases in frequency showed diminishing returns in precision. Thus, we inferred that, for both attack strategies across these two datasets, collecting gradient information ten times from each sample is optimal for distinguishing between member and non-member sets. In other experiments, to strike a balance between efficiency and precision, we will adopt this \textbf{ten-times-per-sample} information collection formula as the default setting.

\paragraph{Different Diffusion Steps and Training Image Resolution.}
In the context of diffusion models, increasing the number of diffusion steps can potentially enhance image quality. This improvement stems from the model's refined capability to capture detailed image nuances by reducing noise over more steps. However, as we add more diffusion steps, the optimization challenge might become more complex. This complexity can slow down the convergence speed and require more detailed hyperparameter adjustments to find the optimal model setup.

When contemplating membership inference attacks, their genesis primarily stems from overfitting during the training phase, leading to discrepancies between the member and non-member samples. We theorize that if adding diffusion steps slows model convergence, it might reduce the overfitting phenomenon, affecting the attack's success. {We set the total diffusion steps from $500$ to $2000$, kept other parameters unchanged, and retrained the model on both ImageNet and CIFAR-10 datasets.}


{In~\autoref{fig:5_a}, we observe that increasing the number of diffusion steps significantly influences our attack success rate, which aligns with our hypothesis. For models trained on CIFAR-10, both $\myattackone$ and $\myattacktwo$ achieve an attack accuracy close to $1.00$ after training with $300$ epochs. However, as the number of denoising steps increases, the attack accuracy decreases by nearly $10\%$ when the denoising step is set to $2000$. The increase in denoising steps leads to a decrease in attack accuracy. This pattern is also observed for models trained on ImageNet when attacked with $\myattackone$ and $\myattacktwo$. We think this is because MIA relies on exploiting the model's overfitting. However, increasing the denoising steps slows down the model’s convergence, thereby impairing the effectiveness of the attack.}

\begin{figure}[t]
    \centering
    \includegraphics[width=0.45\textwidth]{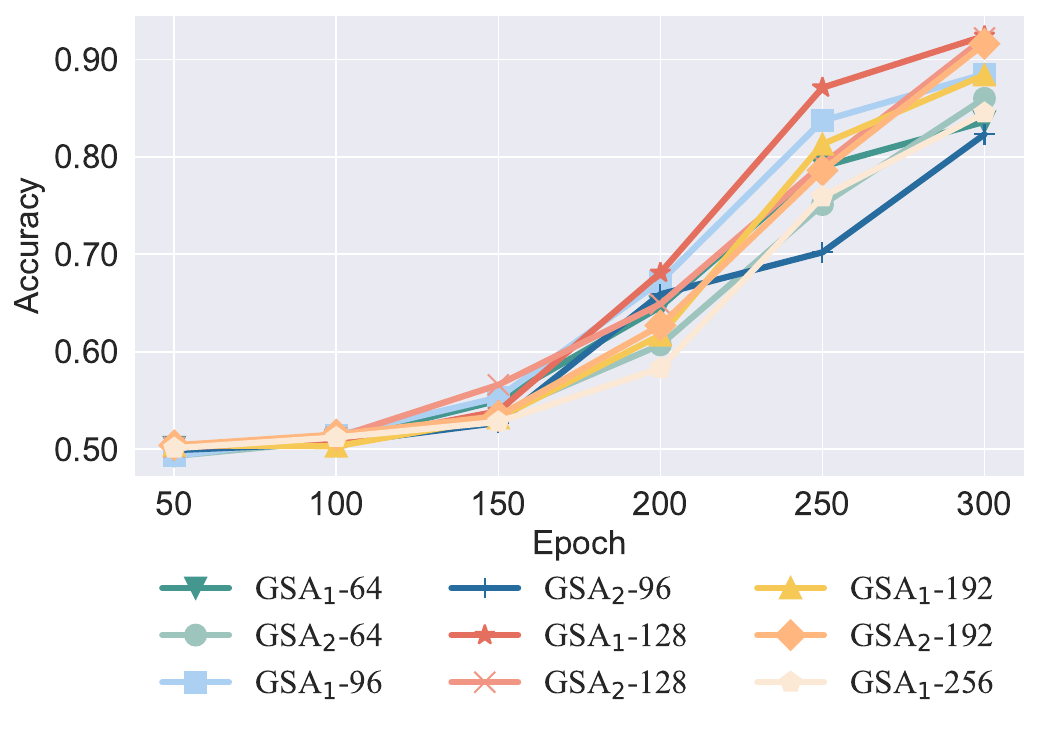}
    \caption{{Results from ImageNet represent the resolution of the image can influence the attack's training accuracy by affecting the model's convergence rate.}}
    \label{resolution}
\end{figure}

Moreover, input data resolution also plays an important role in determining attack success rates. High-resolution images help in distinguishing between member and non-member samples due to their intricate details, but they also require more computational resources and longer training times. Such images may also decelerate the convergence rate of the model, potentially mitigating the extent of overfitting and necessitating additional epochs to achieve equivalent attack outcomes as before.{To investigate the impact of high-resolution images on attack performance, we conducted the experiments using both $\myattackone$ and $\myattacktwo$ on images with resolutions ranging from $64$ to $256$ pixels.}


\begin{table*}[t]
  \centering
  \Huge
  \caption{The table presents the performance results of $\myattackone$ and $\myattacktwo$, trained on three different datasets and evaluated using four distinct evaluation metrics.}
  \label{table:my_result}
  \resizebox{\textwidth}{!}{
\begin{tabular}{ccccccccccccc}
\toprule[2pt]
\multirow{2}{*}{\textbf{\begin{tabular}[c]{@{}c@{}} Attack \\ method\end{tabular}}}& \multicolumn{3}{c}{\textbf{ASR}$^{\uparrow}$} & \multicolumn{3}{c}{\textbf{AUC}$^{\uparrow}$} & \multicolumn{3}{c}{\textbf{TPR@1\%FPR}$^{\uparrow}$}& \multicolumn{3}{c}{\textbf{TPR@0.1\%FPR}$^{\uparrow}$}\\ \cmidrule(lr){2-4} \cmidrule(lr){5-7} \cmidrule(lr){8-10} \cmidrule(lr){11-13} 
 & \begin{tabular}[c]{@{}c@{}}CIFAR-10\end{tabular} & ImagetNet & \begin{tabular}[c]{@{}c@{}}MS COCO \end{tabular} & \begin{tabular}[c]{@{}c@{}}CIFAR-10\end{tabular} & \begin{tabular}[c]{@{}c@{}}ImagetNet\end{tabular} & \begin{tabular}[c]{@{}c@{}}MS COCO \end{tabular} & \begin{tabular}[c]{@{}c@{}}CIFAR-10\end{tabular} & \begin{tabular}[c]{@{}c@{}}ImagetNet\end{tabular} & \begin{tabular}[c]{@{}c@{}}MS COCO \end{tabular} & \begin{tabular}[c]{@{}c@{}}CIFAR-10\end{tabular} & \begin{tabular}[c]{@{}c@{}}ImagetNet\end{tabular} & \begin{tabular}[c]{@{}c@{}}MS COCO \end{tabular} \\ \midrule[0.1pt]
 \multirow{2}{*}{LSA} & \multirow{2}{*}{\begin{tabular}[c]{@{}c@{}}$0.822$\end{tabular}} & \multirow{2}{*}{\begin{tabular}[c]{@{}c@{}}$0.702$\end{tabular}} & \multirow{2}{*}{\begin{tabular}[c]{@{}c@{}}$0.684$\end{tabular}} & \multirow{2}{*}{\begin{tabular}[c]{@{}c@{}}$0.896$\end{tabular}} &\multirow{2}{*}{\begin{tabular}[c]{@{}c@{}}$0.766$\end{tabular}} & \multirow{2}{*}{\begin{tabular}[c]{@{}c@{}}$ 0.746$\end{tabular}}&\multirow{2}{*}{\begin{tabular}[c]{@{}c@{}}$0.146$\end{tabular}} &\multirow{2}{*}{\begin{tabular}[c]{@{}c@{}}$0.034$\end{tabular}}&\multirow{2}{*}{\begin{tabular}[c]{@{}c@{}}$ 0.073$\end{tabular}}&\multirow{2}{*}{\begin{tabular}[c]{@{}c@{}}$0.021$\end{tabular}}&\multirow{2}{*}{\begin{tabular}[c]{@{}c@{}}$0.004$\end{tabular}}&\multirow{2}{*}{\begin{tabular}[c]{@{}c@{}}$0.011$\end{tabular}}\\ \\
\multirow{2}{*}{\myattackone} & \multirow{2}{*}{\begin{tabular}[c]{@{}c@{}}$0.993$\end{tabular}} & \multirow{2}{*}{\begin{tabular}[c]{@{}c@{}}$0.992$\end{tabular}} & \multirow{2}{*}{\begin{tabular}[c]{@{}c@{}}$0.977$\end{tabular}} & \multirow{2}{*}{\begin{tabular}[c]{@{}c@{}}$0.999$\end{tabular}} &\multirow{2}{*}{\begin{tabular}[c]{@{}c@{}}$0.999$\end{tabular}} & \multirow{2}{*}{\begin{tabular}[c]{@{}c@{}}$0.997$\end{tabular}}&\multirow{2}{*}{\begin{tabular}[c]{@{}c@{}}$0.997$\end{tabular}} &\multirow{2}{*}{\begin{tabular}[c]{@{}c@{}}$0.995$\end{tabular}}&\multirow{2}{*}{\begin{tabular}[c]{@{}c@{}}$0.954$\end{tabular}}&\multirow{2}{*}{\begin{tabular}[c]{@{}c@{}}$0.829$\end{tabular}}&\multirow{2}{*}{\begin{tabular}[c]{@{}c@{}}$0.937$\end{tabular}}&\multirow{2}{*}{\begin{tabular}[c]{@{}c@{}}$0.627$\end{tabular}}\\ \\
\multirow{2}{*}{\myattacktwo} &\multirow{2}{*}{\begin{tabular}[c]{@{}c@{}}$0.988$\end{tabular}}& \multirow{2}{*}{\begin{tabular}[c]{@{}c@{}}$0.983$\end{tabular}} & \multirow{2}{*}{\begin{tabular}[c]{@{}c@{}}$0.994$\end{tabular}} & \multirow{2}{*}{\begin{tabular}[c]{@{}c@{}}$0.999$\end{tabular}} & \multirow{2}{*}{\begin{tabular}[c]{@{}c@{}}$0.999$\end{tabular}} & \multirow{2}{*}{\begin{tabular}[c]{@{}c@{}}$0.999$\end{tabular}} & \multirow{2}{*}{\begin{tabular}[c]{@{}c@{}}$0.979$\end{tabular}} & \multirow{2}{*}{\begin{tabular}[c]{@{}c@{}}$0.964$\end{tabular}} & \multirow{2}{*}{\begin{tabular}[c]{@{}c@{}}$0.998$\end{tabular}} & \multirow{2}{*}{\begin{tabular}[c]{@{}c@{}}$0.586$\end{tabular}} & \multirow{2}{*}{\begin{tabular}[c]{@{}c@{}}$0.743$\end{tabular}} & \multirow{2}{*}{\begin{tabular}[c]{@{}c@{}}$0.976$\end{tabular}}\\ \\ \bottomrule[2pt]
\end{tabular}}
\end{table*}

{In~\autoref{resolution}, we observed that the highest attack accuracy was achieved with $\myattackone$ and $\myattacktwo$ when the image resolution was set to $128 \times 128$. The results indicate that lower-resolution samples do not necessarily lead to better attack performance. Increasing the resolution from $64$ to $128$ allows the model to capture more granular details, improving the distinction between member and non-member samples. However, when the resolution is further increased to $256$, a noticeable decline in success rate occurs. We believe this is because higher-resolution images require more training steps for the model to converge. Therefore, when the training time is fixed but the resolution increases, the overfitting phenomenon to the training data diminishes. This reduction in overfitting causes the attack to become less effective. Additionally, both excessively high and low resolutions can negatively impact the final attack performance. An optimal resolution exists where the model can capture sufficient details without requiring extensive training, achieving a balanced fit.}

\begin{tcolorbox}[breakable, colback=takeaways, boxrule=0pt]
\textit{Takeaways:}\:In settings where unconditional diffusion models serve as the target model, overfitting is considered foundational for MIAs. Moreover, distinctions between member and non-member samples can vary at different timesteps. Given these factors, we have investigated several elements that could influence the attack's success rate. These factors encompass the number of training epochs, number of sampling timesteps from a single instance (represented as \( |K| \)), the total diffusion steps, and the resolution of the images. Results from these explorations are presented in the aforementioned figures, with ASR adopted as the evaluation metric.
\end{tcolorbox}

\subsection{Attacking Conditional Diffusion Model}\label{conditional}
\smallskip\noindent
In this section, we design experiments with Imagen, a state-of-the-art generation model in the text-to-image field. We train two shadow models from scratch, using the MS COCO dataset in this part for training purposes. 


\paragraph{Training on Different Epochs.}
In~\autoref{fig:5_b}, consistent with the two attack scenarios posited in~\autoref{unconditional}, we analyze the effect of training steps on the attack success rate for Imagen models. Our categorization is premised on the attacker's knowledge of the target model's training steps. Notably, when the attacker is uncertain about the number of training steps of the target model, we set the training steps of the target model to a fixed value (in this instance, $400,000$ steps). This experimental setup aligns with that of~\autoref{unconditional}.

Consistent with previous experiments using the unconditional diffusion models, a large proportion of the attack success rate for the Imagen model is influenced by the training steps of the target model and shadow models. Precisely, the more the target model overfits the data, the higher the success rate of the MIA, even if the overfitting phenomenon during the shadow model's training is not notably pronounced. For example,~\autoref{fig:5_b} shows that when deploying the $\myattacktwo$ method with the shadow model trained for 200,000 steps, an attack success rate of up to 84.9$\%$ can be achieved if the target model has been trained for 400,000 steps. However, if the target model's training steps are only 200,000, the attack success rate drops to merely 60.7$\%$, representing a nearly 25$\%$ decrease in accuracy. Hence, the degree to which the target model fits the data profoundly influences the effectiveness of the attack. Surprisingly, when the training steps of the shadow models exceed those of the target model, further increasing the training steps for the shadow models leads to a decline in the success rate of MIA attacks. This finding is similar to the phenomenon observed in~\autoref{unconditional} (i.e., the efficacy of the attack is intimately linked to the disparity in data-fitting degrees between the shadow models and their training datasets and the target model with its respective training data.).


\paragraph{Sampling Frequency Variation Analysis.}
It is evident, as depicted in~\autoref{fig:5_c}, that the frequency of information extraction from a single sample by the model plays a pivotal role in influencing the success rate of the attack. Specifically for Imagen, when both shadow models have undergone extensive training iterations, the attack model trained with $\lvert K \rvert = 10$ achieves a remarkable accuracy of $99.4\%$. More intriguingly, when the FPR is controlled at $1\%$ and $0.1\%$, the TPR is recorded at $99.78\%$ and $97.52\%$ respectively. These remarkable findings highlight a substantial increase in accuracy, forming a significant discrepancy compared to the basic accuracy level of $78.1\%$ achieved with $\lvert K \rvert = 1$.

Through the utilization of two approaches, $\myattackone$ and $\myattacktwo$, we seek to elucidate the impact of equidistant timestep sampling frequency on MIA, mainly when applied to large-scale models such as Imagen. The ultimate goal is to ascertain if we can conserve computational resources without compromising attack effectiveness.

In~\autoref{fig:5_c}, we maintain consistent training iterations for both the target and shadow models. This graph depicts how different equidistant timestep sampling frequencies affect the success rate of \myattackone and \myattacktwo. We experimented with four distinct frequencies: $1$, $2$, $5$, and $10$. Evidently, when restricted to one sampling time, the attack success rate plummets to the lowest. When the sampling frequency doubles, the attack success rate sees a notable increase. The outcome difference between two and five sampling times is minimal for \myattackone. Nevertheless, at a frequency of five times, \myattacktwo achieves a success rate comparable to \myattackone with ten sample times. Impressively, ten sampling times boosts \myattacktwo's success rate to nearly 100$\%$, indicating a marked improvement. Given the high accuracy achieved by sampling ten times for each sample, further sampling appears unnecessary.

\begin{tcolorbox}[breakable, colback=takeaways, boxrule=0pt]
\textit{Takeaways:}\: We tested our two attacks primarily on the large-scale model, Imagen, taking into account two factors: the number of training epochs and the timestep sampling frequency. We have examined how overfitting and timestep selection frequency affect the efficacy of our attack strategies. 
\end{tcolorbox}

\begin{table*}[h]
    \vspace{\baselineskip}
    \centering
    \caption{
    {Efficacy of various defensive measures against LSA\textsuperscript{*}, $\myattackone$, and $\myattacktwo$. Specifically, DP-SGD and RandAugment significantly hindered the attacks' effectiveness.}}
    \label{tab:defense_results}
    \resizebox{0.75\textwidth}{!}{
    \Large
    \begin{tabular}{ccccccccccc}
         \toprule[1.3pt]
         \multirow{2}{*}{\textbf{\begin{tabular}[c]{@{}c@{}}   \\ Method\end{tabular}}} & \multicolumn{2}{c}{\textbf{DP-SGD}} & \multicolumn{2}{c}{\textbf{RandAugment}} & \multicolumn{2}{c}{\textbf{RandomHorizontalFlip}} & \multicolumn{2}{c}{\textbf{Cutout}} & \multicolumn{2}{c}{\textbf{No Defense}} \\ \cmidrule(lr){2-3} \cmidrule(lr){4-5} \cmidrule(lr){6-7} \cmidrule(lr){8-9} \cmidrule(lr){10-11} & \textbf{ASR}$^{\uparrow}$ & \textbf{AUC}$^{\uparrow}$ & \textbf{ASR}$^{\uparrow}$ & \textbf{AUC}$^{\uparrow}$ & \textbf{ASR}$^{\uparrow}$ & \textbf{AUC}$^{\uparrow}$ & \textbf{ASR}$^{\uparrow}$ & \textbf{AUC}$^{\uparrow}$ & \textbf{ASR}$^{\uparrow}$ & \textbf{AUC}$^{\uparrow}$ \\ \midrule  LSA\textsuperscript{*} & $0.504$ & $0.508$ & $0.505$ & $0.501$ & $0.524$ & $0.536$ & $0.765$ & $0.846$ & $0.830$ & $0.909$  \\ 
         $\myattackone$ & $0.506$ & $0.511$ & $0.512$ & $0.518$ & $0.793$ & $0.874$ & $0.923$ & $0.977$ & $0.993$ & $0.999$  \\ $\myattacktwo$ & $0.501$ & $0.502$ & $0.504$ & $0.507$ & $0.737$ & $0.811$ & $0.979$ & $0.997$ & $0.988$ & $0.999$ \\ \bottomrule[1.3pt]
    \end{tabular}}
\end{table*}

\section{Ablation Study} \label{Ablation}
Following the framework described in~\autoref{methodology:reduction and aggregation}, our approach effectively subsamples and aggregates gradients across various dimensions. As evident in~\autoref{table:my_result}, both \myattackone and \myattacktwo demonstrate exemplary performance on all experiments. Subsequently, we further explore the potential for subsampling and aggregating information from the model layer dimension. We aim to ascertain how gradient data from the model layer influences the attack success rates of \myattackone and \myattacktwo. Initially, both \myattackone and \myattacktwo extracted gradient information from every layer of the model for the training of the attack model. However, with the increasing size of dataset and growing model complexity, the computational overhead also rises. Thus, we aim to investigate whether it is feasible to ensure the attack success rates of \myattackone and \myattacktwo without necessarily extracting gradient information from all layers of the model.

Pursuant to this idea, We once again conducted experiments using \myattackone and \myattacktwo on datasets, including CIFAR-10, ImageNet, and MS COCO, while maintaining all other settings according to the default configuration in~\autoref{tab:expermients_setup}. We gradually increased the depth of layers from which we collected gradient information. As illustrated in~\autoref{fig:layer}, the x-axis denotes the cumulative number of layers from which gradients are gathered, starting from the top layer. The y-axis employs the True Positive Rate (TPR) at a False Positive Rate (FPR) of $0.1\%$ as the evaluative criterion. The results indicate that as we collect gradient information from increasing layers, the attack success rate correspondingly escalates due to enhanced information accessibility. Remarkably, attaining the highest attack success rate can be achieved merely by gathering gradient data from the top $80\%$ layers of the models. Accordingly, it may not be essential to extract gradient information from each distinct layer of the model, potentially leading to significant computational resource savings.

\begin{figure}[h]
    \centering
    \includegraphics[width=0.47\textwidth]{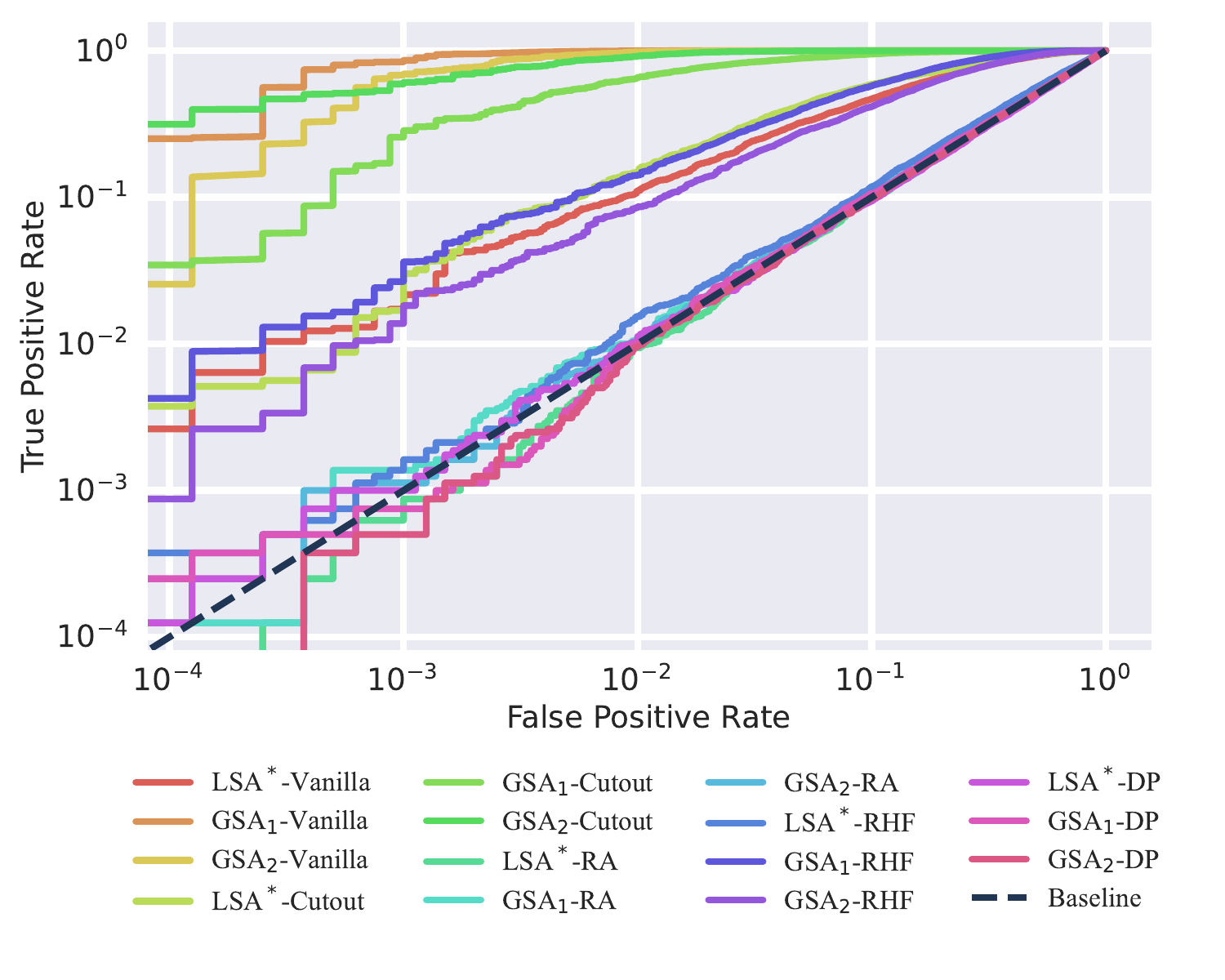}
    \caption{{The performance of LSA\textsuperscript{*}, $\myattackone$ and $\myattacktwo$ under varying defensive strategies is displayed. `Vanilla' refers to the model without any defense methods. `RA' represents RandAugment, and `RHF' denotes RandomHorizontalFlip.}}
    \label{Defense_results}
\end{figure} 

\section{Defenses}\label{defense}

Membership inference attacks are significantly fueled by the overfitting of models to their training data. Thus, mitigating overfitting, such as through data augmentation, could reduce the success rate of these attacks. {We employed various methods of data augmentation~\cite{devries2017improved, cubuk2020randaugment} methods and DP-SGD~\cite{abadi2016deep, dwork2008differential}, a strong privacy-preserving method, as defensive mechanisms against the LSA\textsuperscript{*}, $\myattackone$ and $\myattacktwo$ attacks.} The results following the implementation of these defense mechanisms are presented in~\autoref{tab:defense_results}.


{Firstly, fundamental data augmentation techniques such as Cutout~\cite{devries2017improved} and RandomHorizontalFlip (RHF) were employed as defensive measures. All experiments against LSA\textsuperscript{*}, $\myattackone$, and $\myattacktwo$ were conducted using DDPM~\cite{ho2020denoising} trained on the CIFAR-10 dataset. In these experiments, the model parameters for LSA\textsuperscript{*} were identical to those for $\myattackone$ and $\myattacktwo$, with the only difference being that LSA\textsuperscript{*} used the loss value as attack features. As shown in~\autoref{tab:defense_results}, without any added defense mechanisms, all three attacks achieved high success rates, with $\myattackone$ and $\myattacktwo$ outperforming LSA\textsuperscript{*} (aligned with~\autoref{compare_existing}). When Cutout and RandomHorizontalFlip were applied, LSA\textsuperscript{*} was much more affected than $\myattackone$ and $\myattacktwo$. Specifically, LSA\textsuperscript{*}'s ASR and AUC dropped to around $50\%$ with RHF, while $\myattackone$ and $\myattacktwo$ maintained ASR near $0.80$ and AUC scores are above $0.80$. This represents that when defending against fundamental data augmentations, the gradient-based $\myattackone$ and $\myattacktwo$ are more robust compared to the loss-based LSA\textsuperscript{*}.}



{Then, we evaluated the attack performance of LSA\textsuperscript{*}, $\myattackone$, and $\myattacktwo$ using more powerful defensive strategies: DP-SGD~\cite{abadi2016deep, dwork2008differential} and RandAugment~\cite{cubuk2020randaugment}. DP-SGD, a widely used method, protects training datasets in machine learning by adding noise to the gradient of each sample, thereby ensuring data privacy. In our experiment, we set the clipping bound $C$ to $1$ and the failure probability $\delta$ to $1 \times 10^{-5}$, keeping the experimental settings consistent with the defaults in~\autoref{tab:expermients_setup}. The results show that both DP-SGD and RandAugment effectively defend against LSA\textsuperscript{*} as well as our $\myattackone$ and $\myattacktwo$, reducing the attack ASR and AUC to levels similar to random guessing. The defense effects are also visualized in~\autoref{Defense_results}.
}

\section{Limitation}\label{limitation}
\smallskip\noindent
As shown in~\autoref{table:my_result}, while $\myattackone$ and $\myattacktwo$ can yield satisfactory results with limited computational resources, they are still constrained by their time consumption. Even after implementing subsampling and aggregation across three dimensions, the process of gradient extraction remains time-intensive for larger datasets and more intricate models compared to simply computing the loss. Future studies are anticipated to explore these areas further and identify additional dimensions for reduction. Additionally, the methods employed in this study, $\myattackone$ and $\myattacktwo$, necessitate gradient information from the model for a successful attack. This suggests that requiring complete parameters of the target model during the attack is a rather stringent condition.


\section{Related Work}\label{related_work}
\paragraph{Diffusion Model.} Diffusion model is an emergent generative network originally inspired by diffusion processes from non-equilibrium thermodynamics~\cite{sohl2015deep}. Distinguished from previous Generative Adversarial Networks (GANs)~\cite{creswell2018generative,dhariwal2021diffusion} and Variational Autoencoders (VAEs)~\cite{liang2018variational}, the objective of the diffusion model is to approximate the actual data distribution by engaging a parameterized reverse process that aligns with a simulated diffusion process.

Diffusion models can be connected with score-based models~\cite{song2020score}, generating samples by estimating the gradients of the data distribution and utilizing this gradient information to guide the process of noise addition, thereby producing samples of superior quality. Moreover, the diffusion model showcases the capability to generate images conditioned on specific inputs~\cite{dhariwal2021diffusion, ramesh2022hierarchical, saharia2022palette, meng2021sdedit}.

Apart from generating images, diffusion models are capable of performing specific area retouching in images according to given specifications, hence effectively accomplishing inpainting~\cite{lugmayr2022repaint} tasks. Nowadays, advancements in diffusion models have granted them the ability to generate not only static images but also videos~\cite{ho2022video} and 3D scenes~\cite{gu2023learning}.

\paragraph{Membership Inference Attack.} Membership inference attacks, primarily steered by the seminal work of Homer et al.~\cite{homer2008resolving}, have become an integral part of privacy attack research. The nature of these attacks is typically determined by the depth of information obtained about the target model, whether they are black-box~\cite{choquette2021label,hui2021practical,sablayrolles2019white,salem2018ml,shokri2017membership,song2021systematic,truex2019demystifying,yeom2018privacy} or white-box~\cite{nasr2019comprehensive,rezaei2020towards}. The primary objective lies in determining whether a sample is part of the target model's training set using various metrics functions such as loss~\cite{sablayrolles2019white,yeom2018privacy}, confidence~\cite{salem2018ml}, entropy~\cite{salem2018ml,song2021systematic}, or difficulty calibration~\cite{watson2021importance}. 

\paragraph{Defense.}
 As the popularity of diffusion models continues to rise, a growing body of research quickly unfolds around the privacy and security protections associated with these models. Attacks on diffusion models currently extend beyond mere training data leakage~\cite{carlini2023extracting,duan2023diffusion,fernandez2023privacy,hu2023membership,matsumoto2023membership,wu2022membership} to include the potential use of sensitive data for training~\cite{shan2023glaze}, as well as model theft~\cite{peng2023protecting}. Consequently, effective defense mechanisms against these novel attack types have started to emerge. To prevent the leakage of training data from the target model, privacy distillation~\cite{fernandez2023privacy} methods can be employed. Using this approach, a secure diffusion model can be trained on data generated by the target model after sensitive information has been filtered out. This effectively prevents the leakage of sensitive information during the model training process. For artists concerned about their artwork being used to train diffusion models to generate similar styles, GLAZE~\cite{shan2023glaze} teams suggest adding a watermark to the original art pieces to prevent them from being mimicked by diffusion models. Simultaneously, for every institution, a diffusion model trained using computational resources can be considered one of the company's assets. As such, the desired target model can be fine-tuned to learn a unique diffusion process~\cite{peng2023protecting}, which in turn, contributes to the model's protection.



\section{Conclusion}\label{conclusion}
In this work, we propose a membership inference attack framework that utilizes the norm of the gradient information in diffusion models and presents two specific attack examples, namely $\myattackone$ and $\myattacktwo$. We find that the attack performance on the DDPM and Imagen, trained with the CIFAR-10, ImageNet, and MS COCO datasets, is quite remarkable according to all four evaluation metrics. We posit that a diffusion model's gradient information is more indicative of overfitting to a data point than its loss, hence employing gradient information in MIA could lead to higher success rates. This assertion aligns with the nuanced understanding of model dynamics in the machine learning field. Compared to existing white box loss-based attack methodologies~\cite{hu2023membership,matsumoto2023membership,carlini2023extracting}, our proposed approach demonstrates superior performance under identical model configurations, showcasing efficiency and stability across various datasets and models. This paper introduces the perspective of leveraging gradients for MIA and hopes to inspire valuable follow-up works in this direction.

\section*{Acknowledgment}
We thank the reviewers for their valuable comments and suggestions. This work is partially supported by NSF CNS-2350332, CNS-2350333, OAC-2319988.

\bibliographystyle{acm}
\bibliography{main}

\begin{thebibliography}{10}

\bibitem{abadi2016deep}
{\sc Abadi, M., Chu, A., Goodfellow, I., McMahan, H.~B., Mironov, I., Talwar, K., and Zhang, L.}
\newblock Deep learning with differential privacy.
\newblock In {\em Proceedings of the 2016 ACM SIGSAC conference on computer and communications security\/} (2016), pp.~308--318.

\bibitem{bao2023worth}
{\sc Bao, F., Nie, S., Xue, K., Cao, Y., Li, C., Su, H., and Zhu, J.}
\newblock All are worth words: A vit backbone for diffusion models, 2023.

\bibitem{carlini2022membership}
{\sc Carlini, N., Chien, S., Nasr, M., Song, S., Terzis, A., and Tramer, F.}
\newblock Membership inference attacks from first principles.
\newblock In {\em 2022 IEEE Symposium on Security and Privacy (SP)\/} (2022), IEEE, pp.~1897--1914.

\bibitem{carlini2023extracting}
{\sc Carlini, N., Hayes, J., Nasr, M., Jagielski, M., Sehwag, V., Tramer, F., Balle, B., Ippolito, D., and Wallace, E.}
\newblock Extracting training data from diffusion models.
\newblock {\em arXiv preprint arXiv:2301.13188\/} (2023).

\bibitem{chen2020gan}
{\sc Chen, D., Yu, N., Zhang, Y., and Fritz, M.}
\newblock Gan-leaks: A taxonomy of membership inference attacks against generative models.
\newblock In {\em Proceedings of the 2020 ACM SIGSAC conference on computer and communications security\/} (2020), pp.~343--362.

\bibitem{choquette2021label}
{\sc Choquette-Choo, C.~A., Tramer, F., Carlini, N., and Papernot, N.}
\newblock Label-only membership inference attacks.
\newblock In {\em International conference on machine learning\/} (2021), PMLR, pp.~1964--1974.

\bibitem{creswell2018generative}
{\sc Creswell, A., White, T., Dumoulin, V., Arulkumaran, K., Sengupta, B., and Bharath, A.~A.}
\newblock Generative adversarial networks: An overview.
\newblock {\em IEEE signal processing magazine 35}, 1 (2018), 53--65.

\bibitem{cubuk2020randaugment}
{\sc Cubuk, E.~D., Zoph, B., Shlens, J., and Le, Q.~V.}
\newblock Randaugment: Practical automated data augmentation with a reduced search space.
\newblock In {\em Proceedings of the IEEE/CVF conference on computer vision and pattern recognition workshops\/} (2020), pp.~702--703.

\bibitem{devries2017improved}
{\sc DeVries, T., and Taylor, G.~W.}
\newblock Improved regularization of convolutional neural networks with cutout.
\newblock {\em arXiv preprint arXiv:1708.04552\/} (2017).

\bibitem{dhariwal2021diffusion}
{\sc Dhariwal, P., and Nichol, A.}
\newblock Diffusion models beat gans on image synthesis.
\newblock {\em Advances in Neural Information Processing Systems 34\/} (2021), 8780--8794.

\bibitem{ding2021cogview}
{\sc Ding, M., Yang, Z., Hong, W., Zheng, W., Zhou, C., Yin, D., Lin, J., Zou, X., Shao, Z., Yang, H., et~al.}
\newblock Cogview: Mastering text-to-image generation via transformers.
\newblock {\em Advances in Neural Information Processing Systems 34\/} (2021), 19822--19835.

\bibitem{duan2023diffusion}
{\sc Duan, J., Kong, F., Wang, S., Shi, X., and Xu, K.}
\newblock Are diffusion models vulnerable to membership inference attacks?, 2023.

\bibitem{dwork2008differential}
{\sc Dwork, C.}
\newblock Differential privacy: A survey of results.
\newblock In {\em International conference on theory and applications of models of computation\/} (2008), Springer, pp.~1--19.

\bibitem{fernandez2023privacy}
{\sc Fernandez, V., Sanchez, P., Pinaya, W. H.~L., Jacenków, G., Tsaftaris, S.~A., and Cardoso, J.}
\newblock Privacy distillation: Reducing re-identification risk of multimodal diffusion models, 2023.

\bibitem{ganju2018property}
{\sc Ganju, K., Wang, Q., Yang, W., Gunter, C.~A., and Borisov, N.}
\newblock Property inference attacks on fully connected neural networks using permutation invariant representations.
\newblock In {\em Proceedings of the 2018 ACM SIGSAC conference on computer and communications security\/} (2018), pp.~619--633.

\bibitem{grathwohl2018ffjord}
{\sc Grathwohl, W., Chen, R.~T., Bettencourt, J., Sutskever, I., and Duvenaud, D.}
\newblock Ffjord: Free-form continuous dynamics for scalable reversible generative models.
\newblock {\em arXiv preprint arXiv:1810.01367\/} (2018).

\bibitem{gu2023learning}
{\sc Gu, J., Gao, Q., Zhai, S., Chen, B., Liu, L., and Susskind, J.}
\newblock Learning controllable 3d diffusion models from single-view images.
\newblock {\em arXiv preprint arXiv:2304.06700\/} (2023).

\bibitem{gu2022vector}
{\sc Gu, S., Chen, D., Bao, J., Wen, F., Zhang, B., Chen, D., Yuan, L., and Guo, B.}
\newblock Vector quantized diffusion model for text-to-image synthesis.
\newblock In {\em Proceedings of the IEEE/CVF Conference on Computer Vision and Pattern Recognition\/} (2022), pp.~10696--10706.

\bibitem{hayes2017logan}
{\sc Hayes, J., Melis, L., Danezis, G., and De~Cristofaro, E.}
\newblock Logan: Membership inference attacks against generative models.
\newblock {\em arXiv preprint arXiv:1705.07663\/} (2017).

\bibitem{hilprecht2019monte}
{\sc Hilprecht, B., H{\"a}rterich, M., and Bernau, D.}
\newblock Monte carlo and reconstruction membership inference attacks against generative models.
\newblock {\em Proc. Priv. Enhancing Technol. 2019}, 4 (2019), 232--249.

\bibitem{ho2020denoising}
{\sc Ho, J., Jain, A., and Abbeel, P.}
\newblock Denoising diffusion probabilistic models.
\newblock {\em Advances in Neural Information Processing Systems 33\/} (2020), 6840--6851.

\bibitem{ho2022cascaded}
{\sc Ho, J., Saharia, C., Chan, W., Fleet, D.~J., Norouzi, M., and Salimans, T.}
\newblock Cascaded diffusion models for high fidelity image generation.
\newblock {\em The Journal of Machine Learning Research 23}, 1 (2022), 2249--2281.

\bibitem{ho2022classifierfree}
{\sc Ho, J., and Salimans, T.}
\newblock Classifier-free diffusion guidance, 2022.

\bibitem{ho2022video}
{\sc Ho, J., Salimans, T., Gritsenko, A., Chan, W., Norouzi, M., and Fleet, D.~J.}
\newblock Video diffusion models.
\newblock {\em arXiv preprint arXiv:2204.03458\/} (2022).

\bibitem{homer2008resolving}
{\sc Homer, N., Szelinger, S., Redman, M., Duggan, D., Tembe, W., Muehling, J., Pearson, J.~V., Stephan, D.~A., Nelson, S.~F., and Craig, D.~W.}
\newblock Resolving individuals contributing trace amounts of dna to highly complex mixtures using high-density snp genotyping microarrays.
\newblock {\em PLoS genetics 4}, 8 (2008), e1000167.

\bibitem{hu2021membership}
{\sc Hu, H., and Pang, J.}
\newblock Membership inference attacks against gans by leveraging over-representation regions.
\newblock In {\em Proceedings of the 2021 ACM SIGSAC Conference on Computer and Communications Security\/} (2021), pp.~2387--2389.

\bibitem{hu2023membership}
{\sc Hu, H., and Pang, J.}
\newblock Membership inference of diffusion models.
\newblock {\em arXiv preprint arXiv:2301.09956\/} (2023).

\bibitem{hui2021practical}
{\sc Hui, B., Yang, Y., Yuan, H., Burlina, P., Gong, N.~Z., and Cao, Y.}
\newblock Practical blind membership inference attack via differential comparisons.
\newblock {\em arXiv preprint arXiv:2101.01341\/} (2021).

\bibitem{kim2022diffusionclip}
{\sc Kim, G., Kwon, T., and Ye, J.~C.}
\newblock Diffusionclip: Text-guided diffusion models for robust image manipulation.
\newblock In {\em Proceedings of the IEEE/CVF Conference on Computer Vision and Pattern Recognition\/} (2022), pp.~2426--2435.

\bibitem{kong2023efficient}
{\sc Kong, F., Duan, J., Ma, R., Shen, H., Zhu, X., Shi, X., and Xu, K.}
\newblock An efficient membership inference attack for the diffusion model by proximal initialization.
\newblock {\em arXiv preprint arXiv:2305.18355\/} (2023).

\bibitem{li2021membership}
{\sc Li, J., Li, N., and Ribeiro, B.}
\newblock Membership inference attacks and defenses in classification models.
\newblock In {\em Proceedings of the Eleventh ACM Conference on Data and Application Security and Privacy\/} (2021), pp.~5--16.

\bibitem{liang2018variational}
{\sc Liang, D., Krishnan, R.~G., Hoffman, M.~D., and Jebara, T.}
\newblock Variational autoencoders for collaborative filtering.
\newblock In {\em Proceedings of the 2018 world wide web conference\/} (2018), pp.~689--698.

\bibitem{liu2022membership}
{\sc Liu, Y., Zhao, Z., Backes, M., and Zhang, Y.}
\newblock Membership inference attacks by exploiting loss trajectory.
\newblock In {\em Proceedings of the 2022 ACM SIGSAC Conference on Computer and Communications Security\/} (2022), pp.~2085--2098.

\bibitem{lugmayr2022repaint}
{\sc Lugmayr, A., Danelljan, M., Romero, A., Yu, F., Timofte, R., and Van~Gool, L.}
\newblock Repaint: Inpainting using denoising diffusion probabilistic models.
\newblock In {\em Proceedings of the IEEE/CVF Conference on Computer Vision and Pattern Recognition\/} (2022), pp.~11461--11471.

\bibitem{matsumoto2023membership}
{\sc Matsumoto, T., Miura, T., and Yanai, N.}
\newblock Membership inference attacks against diffusion models, 2023.

\bibitem{meng2021sdedit}
{\sc Meng, C., He, Y., Song, Y., Song, J., Wu, J., Zhu, J.-Y., and Ermon, S.}
\newblock Sdedit: Guided image synthesis and editing with stochastic differential equations.
\newblock In {\em International Conference on Learning Representations\/} (2021).

\bibitem{mukherjee2021privgan}
{\sc Mukherjee, S., Xu, Y., Trivedi, A., Patowary, N., and Ferres, J.~L.}
\newblock privgan: Protecting gans from membership inference attacks at low cost to utility.
\newblock {\em Proc. Priv. Enhancing Technol. 2021}, 3 (2021), 142--163.

\bibitem{nasr2019comprehensive}
{\sc Nasr, M., Shokri, R., and Houmansadr, A.}
\newblock Comprehensive privacy analysis of deep learning: Passive and active white-box inference attacks against centralized and federated learning.
\newblock In {\em 2019 IEEE symposium on security and privacy (SP)\/} (2019), IEEE, pp.~739--753.

\bibitem{nichol2021glide}
{\sc Nichol, A., Dhariwal, P., Ramesh, A., Shyam, P., Mishkin, P., McGrew, B., Sutskever, I., and Chen, M.}
\newblock Glide: Towards photorealistic image generation and editing with text-guided diffusion models.
\newblock {\em arXiv preprint arXiv:2112.10741\/} (2021).

\bibitem{peng2023protecting}
{\sc Peng, S., Chen, Y., Wang, C., and Jia, X.}
\newblock Protecting the intellectual property of diffusion models by the watermark diffusion process, 2023.

\bibitem{raffel2020exploring}
{\sc Raffel, C., Shazeer, N., Roberts, A., Lee, K., Narang, S., Matena, M., Zhou, Y., Li, W., and Liu, P.~J.}
\newblock Exploring the limits of transfer learning with a unified text-to-text transformer.
\newblock {\em The Journal of Machine Learning Research 21}, 1 (2020), 5485--5551.

\bibitem{ramesh2022hierarchical}
{\sc Ramesh, A., Dhariwal, P., Nichol, A., Chu, C., and Chen, M.}
\newblock Hierarchical text-conditional image generation with clip latents, 2022.

\bibitem{ramesh2021zero}
{\sc Ramesh, A., Pavlov, M., Goh, G., Gray, S., Voss, C., Radford, A., Chen, M., and Sutskever, I.}
\newblock Zero-shot text-to-image generation.
\newblock In {\em International Conference on Machine Learning\/} (2021), PMLR, pp.~8821--8831.

\bibitem{rezaei2020towards}
{\sc Rezaei, S., and Liu, X.}
\newblock Towards the infeasibility of membership inference on deep models.
\newblock {\em arXiv preprint arXiv:2005.13702\/} (2020).

\bibitem{rombach2022highresolution}
{\sc Rombach, R., Blattmann, A., Lorenz, D., Esser, P., and Ommer, B.}
\newblock High-resolution image synthesis with latent diffusion models, 2022.

\bibitem{sablayrolles2019white}
{\sc Sablayrolles, A., Douze, M., Schmid, C., Ollivier, Y., and J{\'e}gou, H.}
\newblock White-box vs black-box: Bayes optimal strategies for membership inference.
\newblock In {\em International Conference on Machine Learning\/} (2019), PMLR, pp.~5558--5567.

\bibitem{saharia2022palette}
{\sc Saharia, C., Chan, W., Chang, H., Lee, C., Ho, J., Salimans, T., Fleet, D., and Norouzi, M.}
\newblock Palette: Image-to-image diffusion models.
\newblock In {\em ACM SIGGRAPH 2022 Conference Proceedings\/} (2022), pp.~1--10.

\bibitem{saharia2022photorealistic}
{\sc Saharia, C., Chan, W., Saxena, S., Li, L., Whang, J., Denton, E.~L., Ghasemipour, K., Gontijo~Lopes, R., Karagol~Ayan, B., Salimans, T., et~al.}
\newblock Photorealistic text-to-image diffusion models with deep language understanding.
\newblock {\em Advances in Neural Information Processing Systems 35\/} (2022), 36479--36494.

\bibitem{saharia2022image}
{\sc Saharia, C., Ho, J., Chan, W., Salimans, T., Fleet, D.~J., and Norouzi, M.}
\newblock Image super-resolution via iterative refinement.
\newblock {\em IEEE Transactions on Pattern Analysis and Machine Intelligence 45}, 4 (2022), 4713--4726.

\bibitem{salem2018ml}
{\sc Salem, A., Zhang, Y., Humbert, M., Berrang, P., Fritz, M., and Backes, M.}
\newblock Ml-leaks: Model and data independent membership inference attacks and defenses on machine learning models.
\newblock {\em arXiv preprint arXiv:1806.01246\/} (2018).

\bibitem{shan2023glaze}
{\sc Shan, S., Cryan, J., Wenger, E., Zheng, H., Hanocka, R., and Zhao, B.~Y.}
\newblock Glaze: Protecting artists from style mimicry by text-to-image models, 2023.

\bibitem{shokri2017membership}
{\sc Shokri, R., Stronati, M., Song, C., and Shmatikov, V.}
\newblock Membership inference attacks against machine learning models.
\newblock In {\em 2017 IEEE symposium on security and privacy (SP)\/} (2017), IEEE, pp.~3--18.

\bibitem{sohl2015deep}
{\sc Sohl-Dickstein, J., Weiss, E., Maheswaranathan, N., and Ganguli, S.}
\newblock Deep unsupervised learning using nonequilibrium thermodynamics.
\newblock In {\em International Conference on Machine Learning\/} (2015), PMLR, pp.~2256--2265.

\bibitem{song2020denoising}
{\sc Song, J., Meng, C., and Ermon, S.}
\newblock Denoising diffusion implicit models.
\newblock {\em arXiv preprint arXiv:2010.02502\/} (2020).

\bibitem{song2021systematic}
{\sc Song, L., and Mittal, P.}
\newblock Systematic evaluation of privacy risks of machine learning models.
\newblock In {\em 30th USENIX Security Symposium (USENIX Security 21)\/} (2021), pp.~2615--2632.

\bibitem{song2019generative}
{\sc Song, Y., and Ermon, S.}
\newblock Generative modeling by estimating gradients of the data distribution.
\newblock {\em Advances in neural information processing systems 32\/} (2019).

\bibitem{song2020score}
{\sc Song, Y., Sohl-Dickstein, J., Kingma, D.~P., Kumar, A., Ermon, S., and Poole, B.}
\newblock Score-based generative modeling through stochastic differential equations.
\newblock {\em arXiv preprint arXiv:2011.13456\/} (2020).

\bibitem{truex2019demystifying}
{\sc Truex, S., Liu, L., Gursoy, M.~E., Yu, L., and Wei, W.}
\newblock Demystifying membership inference attacks in machine learning as a service.
\newblock {\em IEEE Transactions on Services Computing 14}, 6 (2019), 2073--2089.

\bibitem{Maaten2008VisualizingDU}
{\sc van~der Maaten, L., and Hinton, G.~E.}
\newblock Visualizing data using t-sne.
\newblock {\em Journal of Machine Learning Research 9\/} (2008), 2579--2605.

\bibitem{von-platen-etal-2022-diffusers}
{\sc von Platen, P., Patil, S., Lozhkov, A., Cuenca, P., Lambert, N., Rasul, K., Davaadorj, M., and Wolf, T.}
\newblock Diffusers: State-of-the-art diffusion models.
\newblock \url{https://github.com/huggingface/diffusers}, 2022.

\bibitem{watson2021importance}
{\sc Watson, L., Guo, C., Cormode, G., and Sablayrolles, A.}
\newblock On the importance of difficulty calibration in membership inference attacks.
\newblock {\em arXiv preprint arXiv:2111.08440\/} (2021).

\bibitem{wu2022membership}
{\sc Wu, Y., Yu, N., Li, Z., Backes, M., and Zhang, Y.}
\newblock Membership inference attacks against text-to-image generation models.
\newblock {\em arXiv preprint arXiv:2210.00968\/} (2022).

\bibitem{ye2022enhanced}
{\sc Ye, J., Maddi, A., Murakonda, S.~K., Bindschaedler, V., and Shokri, R.}
\newblock Enhanced membership inference attacks against machine learning models.
\newblock In {\em Proceedings of the 2022 ACM SIGSAC Conference on Computer and Communications Security\/} (2022), pp.~3093--3106.

\bibitem{yeom2018privacy}
{\sc Yeom, S., Giacomelli, I., Fredrikson, M., and Jha, S.}
\newblock Privacy risk in machine learning: Analyzing the connection to overfitting.
\newblock In {\em 2018 IEEE 31st computer security foundations symposium (CSF)\/} (2018), IEEE, pp.~268--282.

\bibitem{yosinski2014transferable}
{\sc Yosinski, J., Clune, J., Bengio, Y., and Lipson, H.}
\newblock How transferable are features in deep neural networks?
\newblock {\em Advances in neural information processing systems 27\/} (2014).

\bibitem{yu2022scaling}
{\sc Yu, J., Xu, Y., Koh, J.~Y., Luong, T., Baid, G., Wang, Z., Vasudevan, V., Ku, A., Yang, Y., Ayan, B.~K., et~al.}
\newblock Scaling autoregressive models for content-rich text-to-image generation.
\newblock {\em arXiv preprint arXiv:2206.10789\/} (2022).

\end{thebibliography}

\appendix
\section{Additional Information for Denoising Diffusion Probabilistic Model}\label{appendix:DDPM}
The operating mechanism of the diffusion model entails the model learning the posterior probability of the forward process, thereby achieving the denoising process. In the forward noise addition process, assume that there is a sample $x_{t-1}$ at time point $t-1$. Then $x_{t}$ can be represented as:
\begin{align}
    x_t = \sqrt{\alpha_t}x_{t-1}+\sqrt{1-\alpha_t}\epsilon,\;\epsilon\sim \mathcal{N}(0, 1)
    \label{x_t-1}
\end{align} 
Since $\epsilon$ is a random noise, we can unroll the recursive definition and derive $x_t$ directly from $x_0$ (the original image) and time step $t$ (and $\bar{\alpha}_t = \prod_{i=1}^{t} \alpha_i,$):
\begin{align}
    x_t = \sqrt{\bar{\alpha}_t}x_{0}+\sqrt{1-\bar{\alpha}_t}\epsilon_t,\;\epsilon_t\sim \mathcal{N}(0, 1)
    \label{appendix:ddpm_x0}
\end{align} 
The reverse process can be described as:
\begin{align*}
    p_\theta(x_{0:T}) = p(x_{T})\prod_{t=1}^{T} p_\theta(x_{t-1}|x_t)
\end{align*} 
where $x'_{T} \sim \mathcal{N} (0, I)$. The image $x'_{t-1}$ at $t-1$ can be restored from $x'_t$ at time $t$, and can be represented as:  
\begin{align}
    p_\theta(x'_{t-1}|x'_t) = \mathcal{N}(x'_{t-1};{\boldsymbol{\mu}}_\theta(x'_t,t),{\boldsymbol{\Sigma}}_\theta(x'_t,t))
    \label{appendix:p_theta}
\end{align}
In the reverse process, the model aims to use the posterior probability of the forward process to guide the denoising process. 
\begin{align*}
    q(x_{t-1}|x_{t},x_{0}) = \mathcal{N}(x_{t-1};{\boldsymbol{\bar{\mu}}}(x_t,x_0),\boldsymbol{\bar{\beta}_t} \mathbf{I})
\end{align*}
As the $\boldsymbol{\bar{\beta}}_t$ in the posterior probability is also a determined value, the model only needs to learn $\boldsymbol{\bar{\mu}}(x_t,t)$.

In~\autoref{appendix:p_theta}, ${\boldsymbol{\mu}}_\theta(x'_t,t)$ is the predicted mean of the distribution for the sample $x'_{t-1}$ at the preceding timestep, and ${\boldsymbol{\Sigma}}_\theta(x'_t,t)$ denotes the covariance matrix of this distribution. In the original study, ${\boldsymbol{\Sigma}}_\theta (x'_t,t) = \sigma^2_t \mathbf{I}$ is set as untrained time-dependent constants. Consequently, our primary attention is dedicated to the mean ${\boldsymbol{\mu}}_\theta(x'_t,t)$ of the predictive network $p_\theta$. By expanding the aforementioned posterior probability using a probability density function, we can derive the mean and variance of the posterior probability. Given that the variance in $p_\theta(x'_{t-1}|x'_{t})$ is associated with $\beta_t$ and is a deterministic value, our attention is solely on the mean. 

When we express $x_0$ in terms of $x_t$ (from~\autoref{appendix:ddpm_x0}) within the mean $\tilde{\boldsymbol{\mu}}(x_t,x_0)$, the revised $\tilde{\boldsymbol{\mu}}(x_t,x_0)$ then only consists of $x_t$ and random noise $\epsilon_t$. Given that $x_t$ is known at the current time step $t$, the task can be reformulated as predicting the random variable $\epsilon_t$. The ${\boldsymbol{\tilde{\mu}}}(x_t,x_0)$ can be represented as:
\begin{align*}
    \tilde{\boldsymbol{\mu}}(x_t,x_0) = \frac{1}{\sqrt{\alpha_t}}(x_t - \frac{\beta_t}{\sqrt{1-\bar{\alpha}_t}}\epsilon_{t})
\end{align*} 
Concurrently, $\boldsymbol{\mu}_\theta(x'_t,t)$ can be expressed as:
\begin{align*}
    \boldsymbol{\mu}_\theta(x'_t,t) = \frac{1}{\sqrt{\alpha_t}}(x'_t - \frac{\beta_t}{\sqrt{1-\bar{\alpha}_t}}\epsilon_\theta(x'_t,t))
\end{align*} 

Thus, the initial loss function for calculating the prediction of $\boldsymbol{\mu}_\theta(x'_t,t)$
can be reformulated into an equation predicting the noise $\epsilon_\theta(x_t,t)$.
\begin{align}
    &L_t(\theta) \\\nonumber
    =& \E_{x_0,\epsilon}\left[\frac{\beta^2_t}{2\sigma^2_t\alpha_t(1-\alpha_t)}\lVert\epsilon_t-\boldsymbol{\epsilon}_\theta(\sqrt{\bar{\alpha}_t}x_{0}+\sqrt{1-\bar{\alpha}_t}\epsilon_t,t)\lVert^2\right]
    \label{loss:appendix_diffusion}
\end{align} 

It has been observed that DDPM~\cite{ho2020denoising} relies solely on the marginals $q(x_t|x_0)$ during sampling and loss optimization, rather than directly utilizing the joint probability $q(x_{1:T}|x_0)$. Given that many joint distributions share the same marginals, DDIM~\cite{song2020denoising} proposed a non-Markovian forward process as an alternative to the Markovian noise addition process inherent in DDPM. However, the final non-Markovian noise addition is structurally identical to that of DDPM, with the only distinction being the sampling process.
\begin{align*}
    x'_{t-1} = \sqrt{\bar{\alpha}_{t-1}}f_{\theta}(x'_t,t) + \sqrt{1 - \bar{\alpha}_{t-1} - \sigma^2_t} \cdot \epsilon_{\theta}(x'_t,t) + \sigma_t\epsilon
\end{align*}
Where $\alpha_t$ and $\epsilon$ are consistent with the notations used in DDPM. $\sigma_{t}$ represents the variance of the noise. The function 
\begin{align*}
    f_{\theta}(x'_t,t) = \left( \frac{x'_t - \sqrt{1 - \bar{\alpha}_t}\epsilon_{\theta}(x'_t,t)}{\sqrt{\bar{\alpha}_t}} \right)
\end{align*}
denotes the prediction of $x'_0$ at timestep $t$, given $x'_t$ and the pretrained model $\epsilon_\theta$.It is worth noting that when $\sigma_t = 0$, the procedure is referred to as the DDIM sampling process, which deterministically generates a sample from latent variables.

\begin{figure*}
    \centering
    \includegraphics[width=0.9\textwidth]{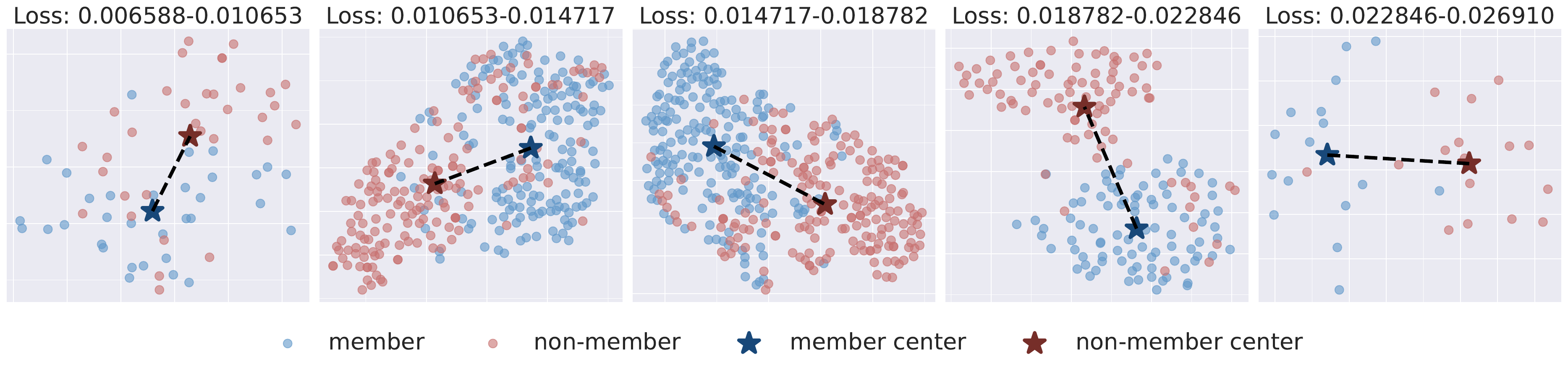}
    \caption{Use t-SNE to represent the member and non-member data pair with the same loss value (rounded to $1e-7$) across five loss intervals. The input to t-SNE is the output of each sample from the last layer of the attack model.}
    \label{fig:all_loss_interval}
\end{figure*}

\section{Additional Likelihood Ratio Attack Details}\label{appendix:carlini}
Carlini et al.~\cite{carlini2022membership} contend that it is erroneous to consider the ramifications of misclassifying a sample as a member of the set as identical to those of incorrect non-member set designation. As a result, they proposed a new evaluation metric and introduced their improved method, LiRA, which proved to be far more effective than previous MIA attack methods in experiments, with up to ten times more efficacy under low False Positive Rates (FPRs). The shadow training technique is also needed here, but it involves creating $\mathbb{D}_{\text{in}}$ and $\mathbb{D}_{\text{out}}$ based on each shadow model's response to the same sample depending on whether the sample was used in the model's training or not. This attack method is white-box as it requires access to the model's output loss and some prior knowledge of the target member's dataset, necessitating the use of target points in the shadow model's training. 

\begin{equation*}
    \label{carlini:likelihood_test}
    \displaystyle \Lambda = \frac{p(\text{conf}_{\text{obs}}\ \mid\ \mathbb{D}_{\text{in}}(x, y))}
    { p(\text{conf}_{\text{obs}}\ \mid\ \mathbb{D}_{\text{out}}(x, y))}
\end{equation*}

The term `$\text{conf}_{\text{obs}}$' refers to the value generated by applying negative exponentiation and logit scaling to the loss produced by the target model for an observed image. `$\mathbb{D}_{\text{in}}$' represents the distribution derived from the processed loss for the member set, while `$\mathbb{D}_{\text{out}}$' stands for the distribution established based on the loss generated for the non-member set samples.

Evidently, the form of LiRA's online attack necessitates retraining the shadow model each time a target point $(x, y)$ is obtained. This approach represents a substantial and arguably uneconomical consumption of resources.

Hence, after proposing this online attack form with many constraints, Carlini et al~\cite{carlini2022membership}. suggested an improved offline attack form that does not require target points in shadow models' training and modifies the attack form to: 
\begin{equation*}
\label{carlini:offline_test}
\Lambda = 1-\Pr[Z > \text{conf}_{\text{obs}}], \text{where } Z \sim \mathbb{D}_{\text{out}}(x, y)) \;.
\end{equation*}

However, the success rate of offline attacks is considerably lower compared to online attacks.

\section{Additional Information for Methodology} \label{appendix:methodology}

In~\autoref{methodology}, we establish the theoretical foundation for \myattackone and \myattacktwo. Specifically, we emphasize that the loss-based attack faces a challenge: \textit{when member and non-member samples have the same loss value, the attack loses effectiveness}. We demonstrate that, in this situation, the gradient data differ between the two samples.

Therefore, we aim to provide experimental evidence to support this claim in this section. Following the attack pipeline, we continue to use gradient data from the shadow model to train an attack model. Then, we compare the loss values of member and non-member samples in the target model. When the loss values of member and non-member samples are the same, we collect them as a data pair. After collecting all data pairs in the target model member/non-member set, we feed all data pairs into the attack model and extract embeddings from the last layer as inputs to do the t-SNE visualization. In~\autoref{fig:all_loss_interval}, we divide the range of loss values into five intervals and present the data pairs in each interval. It is clear that members and non-members can have different gradients in each data pair. Moreover, the member and non-member samples can form distinct clusters. These results indicate that the challenge posed by identical loss values can be overcome by using gradient data, and that gradient data can serve as better features for the attack.

\section{Additional Information for Existing Work}\label{appendix:exasting}

\subsection{SecMI attack}
\label{appendix:secmi}
Drawing from the deterministic reversing and sampling techniques in diffusion models as presented by Song et al.~\cite{song2020score} and Kim et al.~\cite{kim2022diffusionclip}, Duan et al.~\cite{duan2023diffusion} proposed a query-based method that leverages the sampling process and reverse sampling process error at timestep $t$ as the attack feature. The approximated posterior estimation error can be expressed as:
\begin{align*}
    \tilde{\ell}_{t,x_{0}} = \lVert \psi_{\theta}(\phi_{\theta}(\tilde{x}_t,t),t) - \tilde{x}_t \rVert^2
\end{align*}
where 
\begin{equation*}
    \psi_{\theta}(x_t,t) = \sqrt{\bar{\alpha}_{t-1}}f_{\theta}(x_t,t) + \sqrt{1 - \bar{\alpha}_{t-1}}\epsilon_{\theta}(x_t,t)
\end{equation*}
represents the deterministic denoising step, and 
\begin{equation*}
    \phi_{\theta}(x_t,t) = \sqrt{\bar{\alpha}_{t+1}}f_{\theta}(x_t,t) + \sqrt{1 - \bar{\alpha}_{t+1}}\epsilon_{\theta}(x_t,t)
\end{equation*}
signifies the deterministic reverse step(also called DDIM deterministic forward process~\cite{kim2022diffusionclip}) at time $t$, as defined in the original work~\cite{song2020denoising,kim2022diffusionclip,song2020score}.
$\tilde{x}_t$ is obtained from the recursive application of $\phi_{\theta}$, given by $\phi_{\theta}(\ldots\phi_{\theta}(\phi_{\theta}(x_0,0),1),t-1)$. 

Based on $\tilde{\ell}_{t,x_{0}}$, the authors proposed SecMI$_{stat}$ and SecMI$_{NNs}$, which employs the threshold-based attack approach~\cite{yeom2018privacy} and neural network-based attack method~\cite{shokri2017membership}, respectively.

\subsection{Proximal Initialization Attack (PIA)}
\label{appendix:pia}
Building upon the work of Duan et al.~\cite{duan2023diffusion}, Kong et al.~\cite{kong2023efficient} also identified the deterministic properties inherent to the DDIM model~\cite{song2020denoising,song2020score,kim2022diffusionclip}. In the DDIM framework, given $x_0$ and $x_k$, it is feasible to utilize these two points to predict any other ground truth point $x_t$~\cite{kong2023efficient}. Consequently, this methodology employs the $\ell_p$-norm to compute the distance between any ground truth point $x_{t-t'}$ and its predicted counterpart $x{'}_{t-t'}$. After leveraging the ground truth extraction properties of DDIM~\cite{kim2022diffusionclip} and utilizing the sampling formula from~\cite{song2020denoising}, the equation to compute the distance is given by:
\begin{equation*}
    R_{t,p} = \lVert \epsilon_\theta(x_0,0) - \epsilon_{\theta}(\sqrt{\bar{\alpha}_t}x_0 + \sqrt{1 - \bar{\alpha}_t}\epsilon_{\theta}(x_0,0),t) \rVert_p.
\end{equation*}
The notation in the above equation is consistent with the DDPM model, where $R_{t,p}$ denotes the distance. Given that $\epsilon$ is initialized at \( t = 0 \), this method is termed the Proximal Initialization Attack (PIA). When normalizing $\epsilon_{\theta}(x_0,0)$, it is referred to as PIAN (PIA Normalize). This work employs a threshold-based~\cite{yeom2018privacy} attack approach.

Compared to SecMI~\cite{duan2023diffusion}, the attack accuracy has seen a notable improvement. Yet, when juxtaposed with white-box attacks~\cite{hu2023membership,carlini2023extracting}, the success rate of this model attack remains suboptimal. 

\subsection{GAN-Leaks} 
\label{appendix:ganleaks}
GAN-Leaks~\cite{chen2020gan} is a pivotal work in the realm of MIA against GAN models. This work meticulously breaks down attack scenarios into categories based on the level of access to the latent code, generator, and discriminator. For each category, from full black-box to accessible discriminator, GAN-Leaks presents tailored attack methodologies. This work formalizes MIA as an optimization problem. For a given query sample, the goal is to identify the closest reconstruction by optimizing within the generator's output space. A query sample is deemed a member if its reconstruction error is smaller. This can be represented as: 
\begin{equation*}
    \mathcal{R}(x|\mathcal{G}_v) = {G}_v(z^{*}), \text{ where } z^{*} = \underset{z}{\mathrm{argmin}} \ L(x,{G}_v(z))
\end{equation*}
where $L(\cdot,\cdot)$ represents the general distance metric, ${G}_v$ denotes the victim generator, and $z^{*}$ is the optimal estimate.

\begin{figure*}[h]
    \centering
    \includegraphics[width=0.9\textwidth]{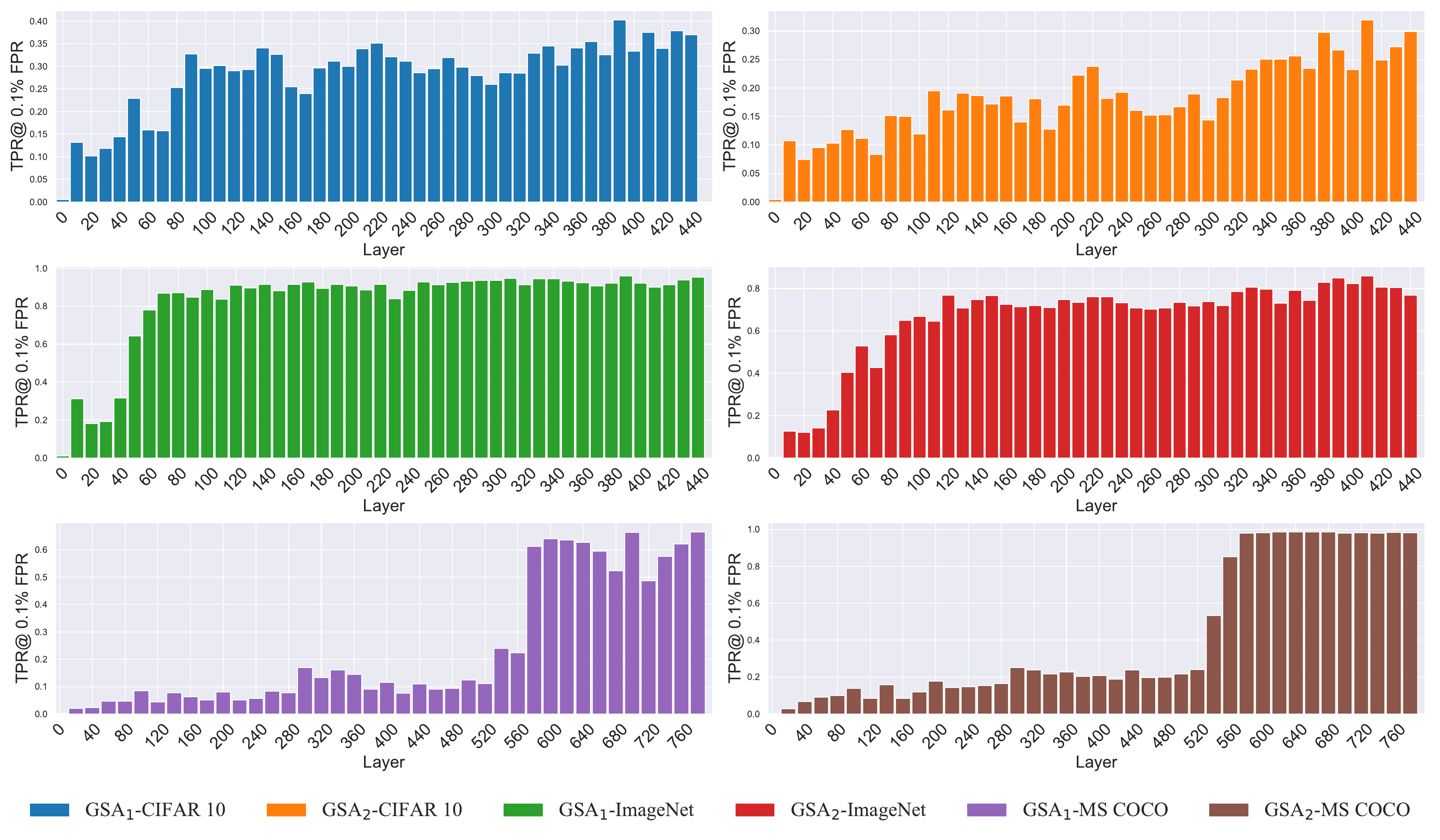}
    \caption{Using \myattackone and \myattacktwo on CIFAR-10, ImageNet, and MS COCO, we can reduce the layers needed for gradient extraction without compromising attack effectiveness. Notably, for attacks on ImageNet-trained DDPM, only $30\%$ of the layers are required for a successful attack.}
    \label{fig:layer}
\end{figure*}

GAN-Leaks~\cite{chen2020gan} is a straightforward attack approach that can be universally applied across diverse settings and generative networks. However, its reliability is contingent upon the quality of the reconstructed image, which can be significantly influenced by the complexity of the original image. A complex image, even if it is from the training set, might encompass intricate details leading to a substantial discrepancy between the reconstructed and query images, resulting in misclassification. To address this, the authors employed a calibration technique to rectify such inaccuracies, ensuring commendable attack accuracy for GAN-Leaks on smaller datasets (comprising fewer than 1000 images). Nonetheless, when applied to extensive datasets, the efficacy of GAN-Leaks diminishes.

\subsection{Likelihood-based Attack}
\label{appendix:likelihood}
The log-likelihood of the samples can be used to conduct a membership inference attack. The formula is given by:
\begin{align*}
\log p(x) = \log p_T(x_T) - \int_{0}^{T} \nabla \cdot \tilde{\mathrm{f}}_{\theta}(x_t,t)\textit{dt}.
\end{align*}

This equation was originally proposed by Song et al.~\cite{song2020score}. If the log-likelihood value exceeds the threshold, the sample is inferred as a member. The term $\nabla \cdot \tilde{\mathrm{f}}_{\theta}(x_t,t)$ is estimated using the Skilling-Hutchinson trace estimator, as suggested by Grathwohl et al.~\cite{grathwohl2018ffjord}.

\section{Additional Information for Ablation Study}
We employed \myattackone and \myattacktwo on CIFAR-10, ImageNet, and MS COCO to further conduct layer-wise reduction as mentioned in~\autoref{methodology:reduction and aggregation}, aiming to reduce computational time and resource consumption. The experimental results are presented in~\autoref{fig:layer}.

\end{document}